\documentclass[12pt]{article}
\usepackage{amsfonts}
\usepackage{amssymb,amsmath}
\usepackage{epsfig}
\usepackage{graphicx}
\usepackage{mdwlist}

\newtheorem{theorem}{Theorem}
\newtheorem{assumption}{Assumption}

\newtheorem{lemma}{Lemma}

\numberwithin{assumption}{section}
\numberwithin{equation}{section}
\numberwithin{theorem}{section}
\numberwithin{lemma}{section}
\numberwithin{figure}{section}
\numberwithin{table}{section}

\begin{document}

\begin{center}
{\large \textbf{Generalized Log-Normal Chain-Ladder}}

\hfill \break
\textsc{D. Kuang}\\[0pt]
\textit{Lloyd's of London, 1 Lime Street, London EC3M 7HA, U.K.}\\[0pt]
{di.kuang@lloyds.com}\\
\textsc{B. Nielsen}\\[0pt]
\textit{Nuffield College, Oxford OX1 1NF, U.K.}\\[0pt]
{bent.nielsen@nuffield.ox.ac.uk}\\[0pt]

\hfill \break
{14 March 2018}
\end{center}

%%%%%%%%%%%%%%%%%%%%%%%%%%%%%%%%%%%%%%%%%%%%%
%%%%%%%%%%%%%%%%%%%%%%%%%%%%%%%%%%%%%%%%%%%%%
%%%%%%%%%%%%%%%%%%%%%%%%%%%%%%%%%%%%%%%%%%%%%
%\newpage
\noindent \textsc{Summary}
We propose an asymptotic theory for distribution forecasting from the log normal chain-ladder model.
The theory overcomes the difficulty of convoluting log normal variables and takes estimation error into account.
The results differ from that of the over-dispersed Poisson model and from the chain-ladder based bootstrap.
We embed the log normal chain-ladder model in a class of infinitely divisible distributions called the
generalized log normal chain-ladder model. The asymptotic theory uses small $\sigma$ asymptotics where
the dimension of the reserving triangle is kept fixed while the standard deviation is assumed to decrease.
The resulting asymptotic forecast distributions follow t distributions.
The theory is supported by simulations and an empirical application.


%%%%%%%%%%%%%%%%%%%%%%%%%%%%%%%%%%%%%%%%%%%%%
%%%%%%%%%%%%%%%%%%%%%%%%%%%%%%%%%%%%%%%%%%%%%
%%%%%%%%%%%%%%%%%%%%%%%%%%%%%%%%%%%%%%%%%%%%%
%\newpage
%\tableofcontents
%\newpage

%%%%%%%%%%%%%%%%%%%%%%%%%%%%%%%%%%%%%%%%%%%%%
%%%%%%%%%%%%%%%%%%%%%%%%%%%%%%%%%%%%%%%%%%%%%
%%%%%%%%%%%%%%%%%%%%%%%%%%%%%%%%%%%%%%%%%%%%%
\section{Introduction} 

Reserving in general insurance usually relies on chain-ladder-type methods.
The most popular method is the traditional chain-ladder. A contender is the log-normal chain-ladder, 
which we study here. Both methods have proved to be valuable for point forecasting.
In practice, distribution forecasting is needed too. For the standard chain-ladder
there are presently three methods available.
Mack (1999)
has suggested a method for recursive calculation of standard errors of the forecasts,
but without proposing an actual forecast distribution.
The bootstrap method of
England and Verrall (1999)
and
England (2002)
is commonly used, but it does not always produce satisfactory results.
Recently,
Harnau and Nielsen (2017)
have developed an asymptotic theory for the chain-ladder in which
the idea of a over-dispersed Poisson framework is embedded in a formal model.
This was done through a class of infinitely divisible distributions and a new
Central Limit Theorem.
An asymptotic theory provides an analytic tool for evaluating the distribution of forecast errors
and building inferential procedures and specification tests for the model.
Here we adapt the infinitely divisible framework of 
Harnau and Nielsen (2017)
to the log-normal chain-ladder
and present an asymptotic theory for the distribution forecasts and model evaluation.
Thereby, asymptotic distribution forecasts and model evaluation tools are now
available for two different models, which together cover a wide range
of reserving triangles. 

The data consists of a reserving triangle of aggregate amounts that have been paid with some
delay in respect to portfolios of insurances.
Table \ref{tab:XLdata}
provides an example.
The objective of reserving is to forecast liabilities that have occurred but have not yet
been settled or even recorded.  The reserve is an estimate of these liabilities. Thus, the problem is to forecast
the lower reserving triangle and then add these forecasts up to get the reserve.
The traditional chain-ladder provides a point forecast for the reserve.

The chain-ladder is maximum likelihood in a Poisson model.  This is useful for estimation
and point forecasting.
Mart{\' i}nez Miranda, Nielsen and Nielsen (2015)
have developed a theory for inference and distribution forecasting in such a Poisson model
in order to analyze and forecast incidences of mesothelioma.
However, this is not of much use for the reserving problem because the
data is nearly always severely over-dispersed.  
The over-dispersion arises because each entry in the paid triangle
is the aggregate amount paid out to an unknown number of claims of different severity.
It is common to interpret this as a compound Poisson variable, see
Beard, Pentik{\" a}inen and Pesonen (1984, \S 3.2).
Compound Poisson variables are indeed over-dispersed in the sense that the variance to mean
ratio is larger than unity. They are, however, difficult to analyze and even harder to
convolute.
England and Verrall (1999)
and
England (2002)
developed a bootstrap to address this issue.  This often works, but it is known to give
unsatisfactory results in some situations.  The model underlying the bootstrap is not fully
described, so it is hard to show formally when the bootstrap is valid and to generalize it to
other situations, including the log-normal chain-ladder.

\setlength{\tabcolsep}{2pt}
\begin{table}[t!]
\centering
%\scriptsize
{\tiny
%\rotatebox{90}
\begin{tabular}{|r|rrrrrrrrrrrrrrrrrrrr|}
                                                                                                                                                                       \hline
     &  1997 &  1998  & 1999  & 2000  & 2001  & 2002  & 2003  & 2004  & 2005  & 2006  & 2007  & 2008  & 2009  & 2010  & 2011  & 2012  & 2013  & 2014  & 2015  & 2016             \\     
                                                                                                                                                                       \hline
1997 &  2185 &  13908 & 44704 & 56445 & 67313 & 62830 & 72619 & 42511 & 32246 & 51257 & 11774 & 21726 & 10926 & 4763  & 3580  & 4777  & 1070  & 1807  & 824   & 1288             \\
1998 &  3004 &  17478 & 49564 & 55090 & 75119 & 66759 & 76212 & 62311 & 31510 & 15483 & 23970 & 8321  & 15027 & 3247  & 8756  & 14364 & 3967  & 3858  & 4643  &                  \\
1999 &  5690 &  28971 & 55352 & 63830 & 71528 & 73549 & 72159 & 37275 & 38797 & 27264 & 28651 & 14102 & 8061  & 17292 & 10850 & 10732 & 4611  & 4608  &       &                  \\
2000 &  9035 &  29666 & 47086 & 41100 & 58533 & 80538 & 70521 & 40192 & 27613 & 13791 & 17738 & 20259 & 12123 & 6473  & 3922  & 3825  & 3082  &       &       &                  \\
2001 &  7924 &  38961 & 41069 & 64760 & 64069 & 61135 & 62109 & 52702 & 36100 & 18648 & 32572 & 17751 & 18347 & 10895 & 2974  & 5828  &       &       &       &                  \\
2002 &  7285 &  25867 & 44375 & 58199 & 61245 & 48661 & 57238 & 29667 & 34557 & 8560  & 12604 & 8683  & 9660  & 4687  & 1889  &       &       &       &       &                  \\
2003 &  3017 &  22966 & 62909 & 54143 & 72216 & 58050 & 29522 & 25245 & 19974 & 16039 & 8083  & 9594  & 3291  & 2016  &       &       &       &       &       &                  \\
2004 &  1752 &  25338 & 56419 & 75381 & 64677 & 58121 & 38339 & 21342 & 14446 & 13459 & 6364  & 6326  & 6185  &       &       &       &       &       &       &                  \\  
2005 &  1181 &  24571 & 66321 & 65515 & 62151 & 43727 & 29785 & 23981 & 12365 & 12704 & 12451 & 8272  &       &       &       &       &       &       &       &                  \\
2006 &  1706 &  13203 & 40759 & 57844 & 48205 & 50461 & 27801 & 21222 & 14449 & 10876 & 8979  &       &       &       &       &       &       &       &       &                  \\
2007 &  623  &  14485 & 27715 & 52243 & 60190 & 45100 & 31092 & 22731 & 19950 & 18016 &       &       &       &       &       &       &       &       &       &                  \\
2008 &  338  &  6254  & 24473 & 32314 & 35698 & 25849 & 30407 & 15335 & 15697 &       &       &       &       &       &       &       &       &       &       &                  \\  
2009 &  255  &  3842  & 14086 & 26177 & 27713 & 15087 & 17085 & 12520 &       &       &       &       &       &       &       &       &       &       &       &                  \\
2010 &  258  &  7426  & 22459 & 28665 & 32847 & 28479 & 24096 &       &       &       &       &       &       &       &       &       &       &       &       &                  \\
2011 &  1139 &  10300 & 19750 & 32722 & 41701 & 29904 &       &       &       &       &       &       &       &       &       &       &       &       &       &                  \\
2012 &  381  &  5671  & 34139 & 33735 & 33191 &       &       &       &       &       &       &       &       &       &       &       &       &       &       &                  \\  
2013 &  605  &  11242 & 24025 & 32777 &       &       &       &       &       &       &       &       &       &       &       &       &       &       &       &                  \\
2014 &  1091 &  9970  & 31410 &       &       &       &       &       &       &       &       &       &       &       &       &       &       &       &       &                  \\
2015 &  1221 &  8374  &       &       &       &       &       &       &       &       &       &       &       &       &       &       &       &       &       &                  \\
2016 &  2458 &        &       &       &       &       &       &       &       &       &       &       &       &       &       &       &       &       &       &                  \\  
                                                                                                                                                                       \hline
\end{tabular}
}
\caption{XL Group, US casualty, gross paid and reported loss and allocated loss adjustment expense in 1000 USD.}
\label{tab:XLdata}
\end{table}
\setlength{\tabcolsep}{6pt}

The infinitely divisible framework of 
Harnau and Nielsen (2017)
provides a plausible over-dispersed Poisson model and framework for distribution forecasting with
the traditional chain-ladder.
It utilizes that the compound Poisson distribution is infinitely divisible.
If the mean of each entry in the paid triangle is large, then the skewness of compound Poisson
variable is small and a Central Limit Theorem applies.
Thus, keeping the dimension of the triangle fixed, while letting the mean increase, the
reserving triangle is asymptotically normal with mean and variance estimated by the chain-ladder.
Since the dimension is fixed we then arrive at an asymptotic theory that matches the
traditional theory for analysis of variance (anova) developed by Fisher in the 1920s.
If the over-dispersion is unity and therefore known as in the Poisson model of
Mart{\' i}nez Miranda, Nielsen and Nielsen (2015)
then inference is
asymptotically $\chi^2$ and distribution forecasts are normal.
When the over-dispersion is estimated as appropriate for reserving data then we arrive
at inference that is 
asymptotically $\mathsf{F}$
and distribution forecasts that are 
asymptotically $\mathsf{t}$.
The chain-ladder bootstrap could potentially be analyzed within this framework, but this is yet
to be done.

When it comes to the log-normal model the situation is different.
The log-normal model has apparently been suggested by
Taylor in 1979,
and then analyzed by for instance
Kremer (1982),
Renshaw (1989),
Verrall (1991, 1994),
Doray (1996)
and
England and Verrall (2002).
The main difference to the over-dispersed Poisson model is that the mean-variance ratio is
constant across the triangle in that model, while the mean-standard deviation ratio is
constant in the log-normal model. Therefore the tails of distributions are expected to be different,
which may matter in distribution forecasting.

Estimation is easy in the log-normal model. It is done by least squares from the log triangle.
Recently,
Kuang, Nielsen and Nielsen (2015)
have provided exact expressions for all estimators along with a set of
associated development factors. 
Least squares theory
provides a distribution theory for the estimators and for inference.
However, the reserving problem is to make forecasts of reserves that are measured on the
original scale. Each entry in the original scale is log-normally distributed. While
there are expressions for such log-normal distributions it is unclear how to incorporate
estimation uncertainty, let alone convolute such variable to get the reserve.

The infinitely divisible theory provides a solution also for the log-normal model.
Thorin (1977) showed that the log-normal distribution is
infinitely divisible.  First of all, this indicates that the log normal variables actually
have an interpretation as compound sums of claims.
Secondly, the framework of
Harnau and Nielsen (2017)
and their Central Limit Theorem apply,
albeit with subtle differences.
In the over-dispersed Poisson model the mean of each entry is taken to be large in the asymptotic
theory, whereas for generalized log-normal model we will let the variance be small in the asymptotic
theory. In both cases the mean-dispersion ratio is then small. 
In this paper we will exploit that infinitely divisible
theory to provide an asymptotic theory for the log-normal
distribution forecasts.

We also discuss specification tests for the log-normal model.
Mis-specification can appear both in the mean and the variance of the log-normal variables.
The mean could for instance have an omitted calendar effect. Thus, we study the
extended chain-ladder model discussed by
Zehnwirth (1994), 
Barnett and Zehnwirth (2000),
and
Kuang, Nielsen and Nielsen (2008a,b,2011).
The variance could be different in subgroups of the triangle as pointed out by
Hertig (1985).
Barlett (1937)
proposed a test for this problem. Recently, Harnau (2017)
has adapted that test to the traditional chain-ladder. We extend this to the
generalized log-normal model.

We illustrate the new methods using a casualty reserving triangle from XL Group (2017)
as shown in
Table \ref{tab:XLdata}.
The triangle is for US casualty and includes gross paid and reported loss and
allocated loss adjustment expense in 1000 USD.

We conduct a simulation study where the data generating process matches the XL Group data in
Table \ref{tab:XLdata}.
We find that that the asymptotic results give good approximations in finite samples.
The asymptotic will work even better if the mean-dispersion ratio is larger.
The generalized log-normal model is also compared with the over-dispersed Poisson model
and the England (2002) bootstrap.  The bootstrap is found not to work very well by an order of 
magnitude for this log-normal data generating process. The over-dispersed Poisson model works
better although it is dominated  by the generalized log-normal model.

In \S\ref{s:logN} we review the well known log-normal models for reserving.
In \S\ref{s:logN:asymp} we set up the asymptotic generalized log-normal model based on the infinitely
divisible framework.  We check that the log-normal model is embedded in this class and show
that the results for inference in the log-normal model caries over to the generalized log-normal
model. We also derive distribution forecasts.
We apply the results to the XL Group data in
\S\ref{s:emp},
while
\S\ref{s:sim} provides the simulation study.
Finally, we discuss directions for future research in \S\ref{s:conclusion}.
All proofs of theorems are provided in an Appendix.

%%%%%%%%%%%%%%%%%%%%%%%%%%%%%%%%%%%%%%%%%%%%%
%%%%%%%%%%%%%%%%%%%%%%%%%%%%%%%%%%%%%%%%%%%%%
%%%%%%%%%%%%%%%%%%%%%%%%%%%%%%%%%%%%%%%%%%%%%
\section{Review of the log-normal chain-ladder model}
\label{s:logN}

A competitor to the chain-ladder is the log-normal model.  In this model
the log of the data is normal so that parameters can be estimated by
ordinary least squares.
%This model has apparently been suggested by
%Taylor in 1979,
%and then analyzed by for instance
%Kremer (1982),
%Renshaw (1989),
%Verrall (1991, 1994),
%Doray (1996),
%and
%Kuang, Nielsen and Nielsen (2015).
We review the log-normal model by describing the structure of the data, the model,
statistical analysis, point forecasts and extension by a calendar effect.

%%%%%%%%%%%%%%%%%%%%%%%%%%%%%%%%%%%%%%%%%%%%%
\subsection{Data}

Consider a standard incremental insurance run-off triangle of dimension $k$.
Each entry is denoted $Y_{ij}$
so that $i$ is the origin year, which can be accident year, policy year or year of account,
while $j$ is the development year.
Collectively we have data
$\mathbf{Y}=\{ Y_{ij} , \forall i,j \in \mathcal{I} \}$,
where
$\mathcal{I}$
is the triangular index set
\begin{equation}
    \mathcal{I}
    =
    \{
        i,j :
        \text{ $i$ and $j$ belong to $(1,\dots ,k)$ with $i+j-1=1,\dots ,k$}
    \}
    .
\end{equation}
Let $n=k(k+1)/2$ be the number of observations in the triangle $\mathcal{I}$.
One could allow more general index sets,
see Kuang, Nielsen and Nielsen (2008a),
for instance to allow for situations where some accidents are fully run-off or only recent calendar
years are available.
We are interested in forecasting the lower triangle with index set
\begin{equation}
    \mathcal{J}
    =
    \{
        i,j :
        \text{ $i$ and $j$ belong to $(1,\dots ,k)$ with $i+j-1=k+1,\dots ,2k-1$}
    \}
    .
\end{equation}

%%%%%%%%%%%%%%%%%%%%%%%%%%%%%%%%%%%%%%%%%%%%%
\subsection{Model}

In the log-normal model the log claims have expectation given by the linear predictor
    \begin{equation}
        \mu _{ij}=\alpha_i+\beta_j+\delta .
        \label{mu}
    \end{equation}
The predictor $\mu_{ij}$ is composed of a
an accident year effect
$\alpha_i$,
a  development year effect
$\beta_j$
and an overall level $\delta$.
The model is then defined as follows.

\begin{assumption}
    \label{as:LN}
    \textbf{log-normal model}.    
    The array $Y_{ij}$, $i,j\in \mathcal{I}$,
    satisfies that
    the variables
    $y_{ij}=\log Y_{ij}$
    are independent normal
    $\mathsf{N}(\mu_{ij}, \omega^2)$
    distributed,
    where the predictor is given by
    (\ref{mu})
\end{assumption}

The parametrisation presented in (\ref{mu}) does not identify
the distribution.
It is common to identify the parameters by setting, for instance,
$\delta=0$
and
$\sum_{j=1}^k \beta_j = 0$.
Such an ad hoc identification makes it difficult to extrapolate the model beyond
the square composed of the upper triangle
$\mathcal{I}$
and the lower triangle
$\mathcal{J}$
and it is not amenable to the subsequent asymptotic analysis. 
Thus, we switch to the canonical parametrisation of 
Kuang, Nielsen and Nielsen (2009, 2015)
so that the model becomes a regular exponential family with freely varying parameters.
The canonical parameter is
\begin{equation}
    \xi =
    \{
        \mu_{11} ,
        \Delta\alpha_2 , \dots , \Delta\alpha_k ,
        \Delta\beta _2 , \dots , \Delta\beta _k 
    \}'
    \label{xi}
    ,
\end{equation}
where
$\Delta \alpha_{i}=\alpha _{i}-\alpha _{i-1}$ is the relative accident year effect and
$\Delta \beta _{j}=\beta _{j}-\beta _{j-1}$ is the relative development year effect, while
$\mu _{11}$
is the overall level.
The length of $\xi$ is denoted
$p$, which is
$p=2k-1$
with the chain-ladder structure.
We can then
write
\begin{equation}
    \mu _{ij}
    =\mu _{11}
    +\sum_{\ell =2}^{i}\Delta \alpha _{\ell }
    +\sum_{\ell =2}^{j}\Delta \beta _{\ell }
    =
    X_{ij}' \xi
    ,
    \label{muxi}
\end{equation}
with the convention that empty sums are zero  and
$X_{ij}\in\mathbb{R}^p$ is the design vector
\begin{equation}
X_{ij}^{\prime }= \{ 1 , 1_{(2\le i)} , \dots , 1_{(k\le i)} , 1_{(2\le j)}
, \dots , 1_{(k\le j)} \} ,  \label{Xij}
\end{equation}
where the indicator function $1_{(m\le i)}$ is
unity if $m\le i$ and zero otherwise.

%%%%%%%%%%%%%%%%%%%%%%%%%%%%%%%%%%%%%%%%%%%%%
\subsection{Statistical analysis}
\label{ss:logN:stats}

The log observations
$y_{ij}=\log Y_{ij}$
have a normal log likelihood given by
\begin{equation}
    \ell_{\log \mathsf{N}} (\xi,\omega^2)
    = -\frac{n}{2} \log (2\pi\omega^2)
    -\frac1{2\omega^2}
    \sum_{i,j\in\mathcal{I}}
    (y_{ij}-X_{ij}'\xi)^2
    .
    \label{log_like_logN}
\end{equation}

Stacking the observations
$y_{ij}=\log Y_{ij}$
and the row vectors
$X_{ij}'$
then gives an observation vector $y$ and a design matrix $X$
and a model equation of the form
\begin{equation}
    y=X\xi+ \varepsilon
    .
    \label{model_logN}
\end{equation}    
The least squares estimator for $\xi$ and the residuals are then
\begin{equation}
    \hat\xi = (X'X)^{-1} X'y,
    \qquad
    \hat\varepsilon_{ij}
    =
    y_{ij}
    -
    X_{ij}'\hat\xi
    .
    \label{est_logN:xi}
\end{equation}
while the variance $\omega^2$ is estimated by
\begin{equation}
    \label{est_logN:omega}
    s^2 = \frac{RSS}{n-p}
    \quad
    \text{where}
    \quad
    RSS
    =
    \sum_{i,j\in\mathcal{I}}
    \hat\varepsilon_{ij}^2
    .
\end{equation}
Kuang, Nielsen and Nielsen (2015)
derive explicit expressions for each coordinate of the canonical parameter
and they provide an interpretation in terms of so-called
geometric development factors.

Standard least squares theory provides a distribution theory for the estimators, see for instance
Hendry and Nielsen (2007),
so that
\begin{equation}
    \hat\xi
    \overset{\mathsf{D}}{=}
    \mathsf{N}
    \{
        \xi
        ,
        \omega^2
        (X'X)^{-1}
    \}
    ,
    \qquad
    s^2
    \overset{\mathsf{D}}{=}
    \chi^2_{n-p}/(n-p)
    .
    \label{est_logN_dist}
\end{equation}
Individual components of $\hat\xi$
will also be normal. Standardizing those components and
replacing $\omega^2$ by the estimate
$s^2$
gives the $\mathsf{t}$-statistic, which is
$\mathsf{t}_{n-p}$ distributed.

We may be interested in testing linear restrictions on $\xi$.
This can be done using $\mathsf{F}$-tests.
For instance, the hypothesis that all
$\Delta\alpha$
parameters are zero would indicate that the policy year effect is constant over time.
Such restrictions can be formulated as
$\xi=H \zeta$
for some known matrix
$H\in\mathbb{R}^{p\times p_H}$
and a parameter vector
$\zeta\in \mathbb{R}^{p_H}$.
In the example of zero
$\Delta\alpha$'s
the $H$ matrix
would select the remaining parameters, the 
$\mu_{11}$ and the $\Delta\beta_j$s.
We then get a restricted design matrix
$X_H=XH$
and a model equation of the form
$y=X_H\zeta + \varepsilon$.
We then get estimators
$$
    \hat\zeta = (X_H'X_H)^{-1} X_H'y,
    \qquad
    s_H^2 = \frac{RSS_H}{n-p_H}
    ,
$$
where the residual sum of squares
$RSS_H
    =
    \sum_{i,j\in\mathcal{I}}
    \hat\varepsilon_{H,ij}^2
$
is formed from the residuals
$
    \hat\varepsilon_{H,ij}
    =
    y_{ij}
    -
    X_{H,ij}'\hat\zeta
$
as before.
The hypothesis can be tested by 
$\mathsf{F}$-statistic
\begin{equation}
    F=
    \frac{\{RSS_H-RSS\}/(p-p_H)}
         {RSS/(n-p)}
    \overset{\mathsf{D}}{=}
    \mathsf{F}(p-p_H,n-p_H)
    .
    \label{F}
\end{equation}

We may also be interested in affine restrictions. 
For instance, the hypothesis that all
$\Delta\alpha$
parameters are known
corresponds the hypothesis of known values of relative ultimates.
This may be of interest in an Bornhuetter-Ferguson context,
see
Margraf and Nielsen (2018).
This is analyzed by restricted least squares which also leads to
$\mathsf{t}$
and
$\mathsf{F}$
statistics.

%%%%%%%%%%%%%%%%%%%%%%%%%%%%%%%%%%%%%%%%%%%%%
\subsection{Point forecasting}

In practice we will want to forecast the variables $Y_{ij}$ on the original scale.
Since $y_{ij}$ is
$\mathsf{N}(\mu_{ij},\omega^2)$
then $Y_{ij}=\exp (y_{ij} )$ is
log-normally distributed with mean
$\exp (\mu_{ij}+\omega^2/2)$.
Thus, the point forecast for the lower triangle $\mathcal{J}$,
as well as the predictor for the upper triangle $\mathcal{I}$,
can be formed as
\begin{equation}
    \tilde Y_{ij}
    =
    \exp (X_{ij}'\hat\xi+\hat\omega^2/2)
    ,
    \label{forecast:logN}
\end{equation}
We will also be interested in distribution forecasting.
However, the log-normal model has the drawback that it is a non-trivial problem to characterize
the joint distribution of the variables on the original scale.
Renshaw (1989)
provides expressions for the covariance matrix of the variables on the original
scale, but a further non-trivial step would be needed to characterize the
joint distribution. Once it comes to distribution forecasting we would also need
to take the estimation error into account. This does not make the problem
easier. We will circumvent these issues by exploiting the infinitely divisible
setup of
Harnau and Nielsen (2017).

%%%%%%%%%%%%%%%%%%%%%%%%%%%%%%%%%%%%%%%%%%%%%
\subsection{Extending with a calendar effect}
\label{ss:logN:apc}

It is common to extend the chain-ladder parametrization with a calendar effect, so that
linear predictor in
(\ref{mu})
becomes
\begin{equation}
    \mu _{ij,apc}
    =\alpha_i+\beta_j+\gamma_{i+j-1}+\delta
    ,
    \label{mu:apc}
\end{equation}
where
$i+j-1$ is the calendar year corresponding to accident year
$i$
and development year
$j$.
This model has been suggesting in insurance by
Zehnwirth (1994).
Similar models have been used in a variety of displines
under the name of age-period-cohort models, where age, period and cohort are our
development, calendar and policy year.
The model has an identification problems. The canonical parameter solution of
Kuang, Nielsen and Nielsen (2008a)
is to write
$
    \mu _{ij,apc}
    =
    X_{ij,apc}' \xi_{apc}
$
where, with $h(i,s)=\max (i-s+1,0)$, we have
\begin{eqnarray}
    \xi_{apc}
    &=&
    (\mu_{11},\nu_a,\nu_c,
        \Delta^2\alpha_3 , \dots , \Delta^2\alpha_k ,
        \Delta^2\beta _3 , \dots , \Delta^2\beta _k ,
        \Delta^2\gamma_3 , \dots , \Delta^2\gamma_k 
    )'
    ,
    \label{xi:apc}
\\
    X_{ij,apc}
    &=&
    \{
        1, i-1 , j-1 ,
        h(i,3), \dots , h(i,k) ,
        h(j,3), \dots , h(j,k) ,
    \notag    
\\
    &&
    \hphantom{\hspace{45mm}}
        h(i+j-1,3), \dots , h(i+j-1,k)         
    \}
    \label{Xij:apc}
    .
\end{eqnarray}
The dimension of these vectors is
$p_{apc}=3k-3$.

This model can be analyzed by the same methods as above.
Stack the design vectors
$X_{ij,apc}'$
to a design matrix
$X_{apc}$
and regress $y$ on $X_{apc}$ to get an
estimator
$\xi_{apc}$
of the form
(\ref{est_logN:xi})
along with a residual sum of squares
$RSS_{apc}$
and a variance estimator
$s^2_{apc}$
The significance of the calendar effect can be tested using an
$\mathsf{F}$-statistic as in
(\ref{F}),
where
$\xi$ and $p$
now correspond to the extended model, while
$\zeta$ and $p_H$
correspond to the chain-ladder specification.

When it comes to forecasting it is necessary to extrapolate the calendar effect.
This has to be done with some care due to identication problem, see
Kuang, Nielsen and Nielsen (2008b, 2011).

\section{The generalized log-normal chain-ladder model}
\label{s:logN:asymp}

The log-normal distribution is infinitely divisible as shown by
Thorin (1977).
We can therefore formulate a class of infinitely divisible distributions
encompassing the log-normal. We will refer to this class of distributions
as the  generalized log-normal chain-ladder model.
In the analysis we 
exploit the setup of
Harnau and Nielsen (2017)
to provide distribution forecasts for the generalized log-normal model.

%%%%%%%%%%%%%%%%%%%%%%%%%%%%%%%%%%%%%%%%%%%%%
\subsection{Assumptions and first properties}

The infinitely divisible setup of
Harnau and Nielsen (2017, \S 3.7)
encompasses the log-normal model.
Recall that a distribution $D$ is infinitely divisible, if for any $m\in\mathbb{N}$,
there are independent, identically distributed random variables
$X_1, \dots ,X_m$
such that
$\sum_{\ell=1}^m X_\ell$
has distribution $D$.
The log-normal distribution is infinitely divisible as shown by
Thorin (1977).
This matches the fact that the paid amounts are aggregates of number of payments.
In our data analysis we neither know the number nor the severities of the payments.
Due to the infinite divisibility the log-normal distribution can therefore be a good choice
for modelling aggregate payments.

We will need two assumptions. The first assumption is about a general
infinite divisible setup. The second assumption gives more specific details on the log-normal setup.
\begin{assumption}
    \label{as:HN:inf_div}
    \textbf{Infinite divisibility}.
    The array $Y_{ij}$, $i,j\in \mathcal{I}$,
    satisfies

    \noindent
    $(i)$ $Y_{ij}$ are independent distributed, non-negative and infinitely divisible;
    
    \noindent
    $(ii)$ asymptotically, the dimension of the array $\mathcal{I}$ is fixed;

    \noindent
    $(iii)$ asymptotically, the skewness vanishes:
    $
    \mathsf{skew} (Y_{ij})
    =
    \mathsf{E}
    [
        \{
            Y_{ij} - \mathsf{E} (Y_{ij})
        \}
        /
        \mathsf{sdv} (Y_{ij)}
    ]^3
    \to 0.
    $
\end{assumption}    

We have the following Central Limit Theorem for non-negative, infinitely divisible distributions
with vanishing skewness.  This is different from the standard Lindeberg-L{\'e}vy
Central Limit Theorem for averages of independent, identically distributed variables, but proved
in a similar fashion by analyzing characteristic function and exploiting the L{\'e}vy-Kintchine
formula for infinitely
divisible distributions.
\begin{theorem}
    \label{t:HN:CLT}
    \textbf{(Harnau and Nielsen, 2017, Theorem 1)}
    Suppose Assumption
    \ref{as:HN:inf_div}
    is satisfied. Then
    $$
        \frac{Y_{ij}-\mathsf{E}(Y_{ij})}
             {\surd \mathsf{Var}(Y_{ij})}
        \overset{\mathsf{D}}{\to}
        \mathsf{N} (0,1)
        .
    $$
\end{theorem}

We need some more specific assumptions for the log-normal setup.
%Assumption \ref{as:HN:inf_div}
%will be needed again, while
%the over-dispersed Poisson Assumption \ref{as:HN:inf_div:ODP}
%will be replaced by the following assumption.
When describing the predictor we write $\mu_{ij}=X_{ij}'\xi$
to indicate that any linear structure is allowed as long as $\xi$ is freely varying when
estimating in the statistical model.
This could be the
chain-ladder structure as in
(\ref{muxi}),
(\ref{Xij})
or an extended chain-ladder model with a calendar effect. 

\begin{assumption}
    \label{as:HN:inf_div:LN}
    \textbf{The generalized log-normal chain-ladder model}.    
    The array $Y_{ij}$, $i,j\in \mathcal{I}$,
    satisfies Assumptions \ref{as:HN:inf_div} and the following:

    \noindent
    $(i)$ 
    $
        \log
        \mathsf{E}Y_{ij}=\mu_{ij}+\omega^2/2=X_{ij}'\xi+\omega^2/2 ,
    $
    where $\xi$ is identified by the likelihood
    (\ref{log_like_logN});
    
    \noindent
    $(ii)$
    asymptotically,
    $\omega^2\to 0$
    while $\xi$ is fixed;

    \noindent
    $(iii)$
    asymptotically,
    $\mathsf{Var}(Y_{ij})/\{ \omega^2\mathsf{E}^2(Y_{ij})\}\to 1$.
\end{assumption}    

We check that the log-normal model set out in
Assumption \ref{as:LN}
is indeed of the generalized log-normal model.

\begin{theorem}
    \label{t:log_normal_distribution}
    Consider the log-normal model of 
    Assumption \ref{as:LN}.
    Suppose the dimension of the array
    $\mathcal{I}$ is fixed as   
    $\omega^2\to 0$.
    Then 
    Assumptions \ref{as:HN:inf_div}, \ref{as:HN:inf_div:LN}
    are satisfied.
\end{theorem}

A first consequence of the generalized log-normal model is that
Theorem \ref{t:HN:CLT}
provides an asymptotic theory for the claims on the original scale.
We now check that we have a normal theory for the log claims.
The proof applies the delta method. Theorem \ref{t:asymp_logY} is useful in 
deriving the inference in Theorem \ref{t:asymp_est} and estimation 
error for forecasts in Theorem Theorem \ref{t:asymp_forecast} in later sections.
\begin{theorem}
    \label{t:asymp_logY}
    Suppose 
    Assumptions \ref{as:HN:inf_div}, \ref{as:HN:inf_div:LN}
    are satisfied.
    Let
    $y_{ij}=\log Y_{ij}$.
    Then, as $\omega^2\to 0$,
    $$
        \omega^{-1}
        (   y_{ij} - \mu_{ij} )
        \overset{\mathsf{D}}{\to}
        \mathsf{N} (0,1)
        .
    $$
    Due to the independence of $Y_{ij}$ over $i,j\in\mathcal{I}$
    then the standardized $y_{ij}$ are asymptotically independent
    standard normal.
\end{theorem}

We will need to reformulate the Central Limit Theorem
\ref{t:HN:CLT}
slightly.
The issue is that the generalized log-normal model leaves
the variance of the variable unspecified in finite sample,
so that the Central Limit Theorem is difficult to manipulate directly.  
Theorem \ref{t:asymp_Y} is useful in 
deriving the process error for forecasts in Theorem \ref{t:asymp_forecast} later.
\begin{theorem}
    \label{t:asymp_Y}
    Suppose 
    Assumptions \ref{as:HN:inf_div}, \ref{as:HN:inf_div:LN}
    are satisfied.
    Then, as $\omega^2\to 0$,
    $$
        \omega^{-1}
        \{
            Y_{ij} -
            \mathsf{E} (Y_{ij} )
        \}    
        \overset{\mathsf{D}}{\to}
        \mathsf{N} \{ 0, \exp (2\mu_{ij} ) \}
        .
    $$
    Note that $Y_{ij}$ over $i,j\in\mathcal{I}$
    are assumed independent.
\end{theorem}

%%%%%%%%%%%%%%%%%%%%%%%%%%%%%%%%%%%%%%%%%%%%%
\subsection{Inference}
\label{ss:logN:asymp:inf}

We check that the inferential results for the log-normal model, described in
\S\ref{ss:logN:stats},
carry over to the generalized log-normal model.  First, we consider the asymptotic distribution
of estimators and then the properties of $\mathsf{F}$-statistics for inference. 

\begin{theorem}
    \label{t:asymp_est}
    Consider the generalized log-normal model defined by Assumptions
    \ref{as:HN:inf_div},
    \ref{as:HN:inf_div:LN}
    and the least squares estimators
    (\ref{est_logN:xi}).
    Then, as
    $\omega^2\to 0$,
    \begin{eqnarray*}
        \omega^{-1}(\hat\xi - \xi) &\xrightarrow{\mathsf{D}}&
        \mathsf{N} \{ 0,  (X^{'}X)^{-1} \}
        ,
    \\
        \omega^{-2}s^2
        &\xrightarrow{\mathsf{D}}&
        \chi^2_{n-p}/(n-p)
        .
    \end{eqnarray*}
    The estimators
    $\hat\xi$
    and
    $s^2$
    convergence jointly and are asymptotically independent.
\label{th:infest}
\end{theorem}

We can derive inference for of the estimator $\hat\xi$ using 
asymptotic $\mathsf{t}$  distribution. The proof follows Theorem \ref{th:infest} 
and the Continuous Mapping Theorem. 
\begin{theorem}
\label{t:asymp:inf:t}
 Consider the generalized log-normal model, defined by Assumptions
    \ref{as:HN:inf_div},
    \ref{as:HN:inf_div:LN}. 
Then as $\omega^2\to 0$,
\begin{eqnarray*}
\frac{v' (\hat{\xi} - \xi)}{s\sqrt{v'(X'X)^{-1}}v}\xrightarrow{\mathsf{D}} \mathsf{t}_{n-p}
\end{eqnarray*}
\end{theorem}

We can also make inference using asymptotic
$\mathsf{F}$-statistics,
mirroring the 
$\mathsf{F}$-statistic
(\ref{F})
from the classical normal model. The proof is similar
to Theorem 4 of Harnau and Nielsen (2017).
\begin{theorem}
\label{t:asymp:inf}
    Consider the generalized log-normal model, defined by Assumptions
    \ref{as:HN:inf_div},
    \ref{as:HN:inf_div:LN}
    with three types of linear predictor:

    the extended chain-ladder model parametrised by
    $\xi_{apc}\in\mathbb{R}^{p_{apc}}$
    in
    (\ref{xi:apc});
    
    the chain-ladder model parametrised by
    $\xi\in\mathbb{R}^{p}$
    in
    (\ref{xi}); and

    a linear hypothesis
    $\xi = H\zeta$ for
    $\zeta\in\mathbb{R}^{p_H}$
    and
    some known matrix
    $H\in\mathbb{R}^{p\times p_H}$.

    \noindent
    Let
    $RSS_{apc}$,
    $RSS$
    and
    $RSS_H$
    be the residual sums of squares under the linear hypotheses.
    Then, as $\omega\to 0$,
    \begin{eqnarray*}
        F_1
        &=&
        \frac{(RSS-RSS_{apc})/(p_{apc}-p)}
             {RSS_{apc}/(n-p_{apc})}
        \overset{\mathsf{D}}{\to}
        \mathsf{F}_{p-p_{apc},n-p_{apc}},
        \\
        F_2
        &=&
        \frac{(RSS_H-RSS)/(p-p_H)}
             {RSS/(n-p)}
        \overset{\mathsf{D}}{\to}
        \mathsf{F}_{p_H-p,n-p},
    \end{eqnarray*}
    where
    $F_1$
    and
    $F_2$
    are asymptotically independent. 
\end{theorem}

%%%%%%%%%%%%%%%%%%%%%%%%%%%%%%%%%%%%%%%%%%%%%
\subsection{Distribution forecasting}
\label{ss:logN:asymp:forecast}

The aim is to predict a sum of elements in the lower triangle, that could be the overall
sum, which is the total reserve; or it could be
row sums or diagonal sums giving a cash flow. 
We denote such sums by
${Y}_{\mathcal{A}}
    = \sum_{(i,j)\in\mathcal{A}} {Y}_{ij}$
for some
subset $\mathcal{A}\in\mathcal{J}$.
The point forecasts for a single entry are
$\hat Y_{ij}=\exp (X_{ij}'\hat\xi + s^2/2)$
as given in
(\ref{forecast:logN}), while the overall point forecast is
\begin{equation}
    \tilde{Y}_{\mathcal{A}}
    = \sum_{(i,j)\in\mathcal{A}}
    \tilde{Y}_{ij}
    = \sum_{(i,j)\in\mathcal{A}}
    \exp (X_{ij}'\hat\xi + s^2/2)    
\end{equation}

To find the forecast error we expand
\begin{multline}
    Y_{ij}
    -
    \tilde Y_{ij}
    =
    \{
        Y_{ij}
        -
        \mathsf{E} (Y_{ij})
    \}
    +
    \exp (\omega^2/2)
    \{
        \exp (X_{ij}'\hat\xi)
        -
        \exp (X_{ij}' \xi )
    \}
\\    
    +
    \{
        \exp (\omega^2/2)
        -
        \exp (s^2/2)    
    \}
        \exp (X_{ij}' \xi )
    ,
    \label{forecast:logN:taxonomy}
\end{multline}
which we will sum over $\mathcal{A}$.
This is sometimes called the forecast taxonomy.
This expansion gives some insight into the
asymptotic forecast distribution, although the 
detailed proof will be left to the appendix. 
The first term in
(\ref{forecast:logN:taxonomy})
is the process error. When extending
Theorem \ref{t:asymp_Y}
to the lower triangle $\mathcal{J}$ we will get
\begin{equation}
    \omega^{-1}
    \{
        Y_{\mathcal{A}}
        -
        \mathsf{E} (Y_{\mathcal{A}})
    \}
    \overset{\mathsf{D}}{\to}
    \mathsf{N} (0, \varsigma^2_{\mathcal{A},process})
    ,
    \label{forecast:error_process}
\end{equation}
where
\begin{equation}
    \varsigma^2_{\mathcal{A},process}
    =
    \sum_{i,j\in\mathcal{A}}
    \exp (2 X_{ij}' \xi )
    \label{forecast:var_process}
\end{equation}
The second term in
(\ref{forecast:logN:taxonomy})
is the estimation error for the canonical parameter $\xi$.
From Theorem \ref{t:asymp_est}
we will be able to derive
\begin{equation}
    \omega^{-1}
    \exp (\omega^2/2)
    \{
        \exp (X_{ij}'\hat\xi)
        -
        \exp (X_{ij}' \xi )
    \}
    \overset{\mathsf{D}}{\to}
    \mathsf{N} (0, \varsigma^2_{\mathcal{A},estimation})
    ,
    \label{forecast:error_estimation}
\end{equation}
where
\begin{equation}
    \varsigma^2_{\mathcal{A},estimation}
    =
    \{
    \sum_{i,j\in\mathcal{A}}
    \exp (X_{ij}' \xi )
    X_{ij}'
    \}
    (X'X)^{-1}
    \{
    \sum_{i,j\in\mathcal{A}}
    \exp (X_{ij}' \xi )
    X_{ij}
    \}
    .
    \label{forecast:var_estimation}
\end{equation}
The third term in
(\ref{forecast:logN:taxonomy})
vanishes asymptotically.
We will estimate $\omega^2$ by
$s^2$,
which turns the asymptotic normal distributions into $\mathsf{t}$-distribution.
The process error and the estimation error are asymptotically independent as they are based on
independent variables for the upper and lower triangle,
$\mathcal{J}$
and
$\mathcal{I}$.
We can describe the asymptotic forecast error as follows.
\begin{theorem}
    \label{t:asymp_forecast}
    Suppose the generalized log-normal model defined by Assumptions
    \ref{as:HN:inf_div},
    \ref{as:HN:inf_div:LN}
    applies both in the upper and the lower triangle,
    $\mathcal{I}$
    and
    $\mathcal{J}$.
    Then, as
    $\omega^2\to 0$,
\begin{eqnarray*}
    \label{th:forecastdist}
    \hat\omega^{-1}
    (Y_\mathcal{A} - \tilde{Y}_\mathcal{A})
    \overset{\mathsf{D}}{\to}
    (
        \varsigma_{\mathcal{A},process}^2
        +
        \varsigma_{\mathcal{A},estimation}^2
    )^{1/2}
    \mathsf{t}_{n-p}
    ,
\end{eqnarray*}
where
$\varsigma_{\mathcal{A},process}^2$
and
$\varsigma_{\mathcal{A},estimation}^2$
can be estimated consistently by
\begin{eqnarray}
    r^2_{\mathcal{A},process}
    &=&
    \sum_{i,j\in\mathcal{A}}
    \exp (2 X_{ij}' \hat\xi )
    ,
\\
    r^2_{\mathcal{A},estimation}
    &=&
    \{
    \sum_{i,j\in\mathcal{A}}
    \exp (X_{ij}' \hat\xi )
    X_{ij}'
    \}
    (X'X)^{-1}
    \{
    \sum_{i,j\in\mathcal{A}}
    \exp (X_{ij}' \hat\xi )
    X_{ij}
    \}
    .
\end{eqnarray}
\end{theorem}

Thus, the distribution forecast is
\begin{equation}
    \tilde{Y}_\mathcal{A}
    + \{
        \hat\omega^2
        (
            r_{\mathcal{A},process}^2
            +
            r_{\mathcal{A},estimation}^2
        )
    \}^{1/2}
    \mathsf{t}_{n-q}
    .
\label{LN_distribution_forecast}
\end{equation}

%%%%%%%%%%%%%%%%%%%%%%%%%%%%%%%%%%%%%%%%%%%%%
\subsection{Specification test}
\label{ss:logN:asymp:spec}

Specification tests for the log-normal model can be carried out by allowing a richer structure
for the predictor or for the variance. We have already seen how the generalized log-normal chain-ladder model
can be tested against the extended chain-ladder model using an asymptotic
$\mathsf{F}$-test.
We can test whether the variance is constant across the upper triangle by adopting the
Bartlett (1937) test.
Recently,
Harnau (2017)
has shown how to do model specification tests for the over-dispersed Poisson model. Here we will adapt the Bartlett test to the log-normal chain-ladder.
It should be noted that one can of course also allow a richer structure for the predictor
and the variance simultaneously following the principles outlined here.

\begin{figure}[ht]
    \centering
    \includegraphics[scale=0.7]{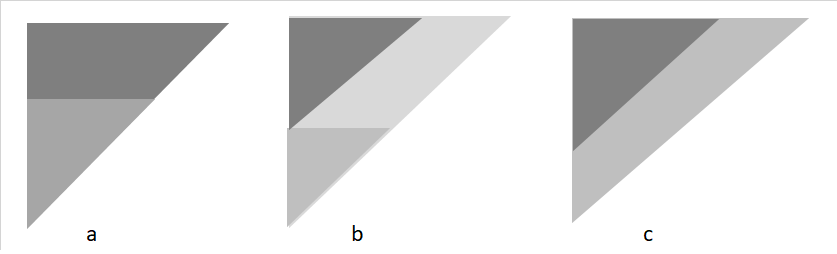}
\caption{Examples of dividing triangles in two parts}
\label{F:Test_Triangles_cut}
\end{figure}

Suppose the triangle $\mathcal{I}$ can be divided into two or more groups as indicated in
Figure \ref{F:Test_Triangles_cut}.
Thus, the index set $\mathcal{I}$ is divided into disjoint sets
$\mathcal{I}_\ell$
for $\ell=1,\dots ,m$.
We then set up a log-normal chain-ladder seperately for each group.
Note that the full canonical parameter vector $\xi$ may not be identified on the subsets.
As we will only be interested in the fit of the models we can ad hoc identify $\xi$ by dropping
sufficiently many columns of the design matrix $X$. This gives us a parameter $\xi_\ell$ and a design vector
$X_{ij\ell}$ for each subset 
$\mathcal{I}_\ell$
and a predictor
$\mu_{ij\ell}=X_{ij\ell}'\xi_\ell$.
Thus the model for each group is that
$y_{ij\ell}$
is
$\mathsf{N}(\mu_{ij\ell},\omega_\ell^2)$.
Let $p_\ell$ denote the dimension of these vectors, while $n_\ell$ is the number of
elements in 
$\mathcal{I}_\ell$ giving
the degrees of freedom $df_\ell=n_\ell-p_\ell$.

When fitting the log-normal chain-ladder seperately to each group we get estimators $\hat\xi_\ell$
and predictors
$\hat\mu_{ij\ell}=X_{ij\ell}'\hat\xi_\ell$.
From this we can compute the residual sum of squares and variance estimators as
\begin{equation}
    RSS_\ell =
    \sum_{i,j\in\mathcal{I}_\ell}
    (
        y_{ij} - \hat\mu_{ij,\ell}
    )^2
    ,
    \qquad
    s^2_\ell
    =
    \frac1{df_\ell} RSS_\ell
    .
\end{equation}

If there are only two subsets then we have two choices of tests available.
The first test is a simple
$\mathsf{F}$-test
for the hypothesis that
$\omega_1=\omega_2$.
In the log-normal model this is
\begin{equation}
    F^\omega
    =
    s_2^2/s_1^2
    \overset{\mathsf{D}}{=}
    \mathsf{F}_{n_2-p_2,n_1-p_1}
    .
\end{equation}
In the generalized log-normal the $\mathsf{F}$-distribution can be shown to be valid
asymptotically. Harnau (2017) has proved this for the over-dispersed Poisson model using
an infinitely divisible setup. That proof extends to the generalized log-normal setup following
the ideas of the proofs of the above theorems. 
We can then construct a two sided test. Choosing a 5\% level this test rejects when
$F^\omega$ is either smaller than the 2.5\% quantile or larger than the 97.5\% quantile of the
$\mathsf{F}_{n_2-p_2,n_1-p_1}$-distribution.

The second test is known as Bartlett's test and applies to any number of groups.
Thus, suppose we have $m$ groups and want to test
$\omega_1=\cdots =\omega_m$.
In the exact log-normal case then
$s_1^2,\dots ,s_m^2$
are independent scaled $\chi^2$ variables.
Bartlett found the likelihood for this $\chi^2$ model.
Under the hypothesis the common variance is estimated by
\begin{equation}
    \bar{s}^2
    =
    \frac1{df_\cdot}
    \sum_{\ell=1}^m RSS_\ell
    ,
    \qquad
    \text{where}
    \qquad
    df_\cdot
    =
    \sum_{\ell=1}^m df_\ell
    =
    n - \sum_{\ell=1}^m p_\ell
    ,
\end{equation}
while the likelihood ratio test statistic for the hypothesis is
\begin{equation}
    LR^\omega
    =
    df_\cdot \log (\bar{s}^2)
    -
    \sum_{\ell=1}^m
    df_\ell \log (s_\ell^2)
%    \overset{\mathsf{D}}{=}
%    \mathsf{Ba}(df_1, \dots , Ba_m)
    .
    \label{Bartlett_test}
\end{equation}
The exact distribution of the likelihood ratio test statistic
depends on the degrees of freedom of the groups, but not on their ordering.
No analytic expression is known.
However, Bartlett showed that this distribution
is very well approximated by a scaled $\chi^2$-distribution. That is
\begin{equation}
    \frac{LR^\omega}{C}
    \approx
    \chi^2_{m-1}
    \qquad
    \text{where}
    \qquad
    C = 1 + \frac1{3(m-1)}
    (
        \sum_{\ell=1}^m
        \frac1{df_\ell}
        -
        \frac1{df_\cdot}
    )
    .
    \label{Bartlett}
\end{equation}
The factor $C$ is known as the Bartlett correction factor.  Formally, the approximation
is a second order expansion which is valid when the small group is large, so that
$\min_\ell df_\ell$ is large. However, the approximation works exceptionelly well in very small samples;
see the simulations by Harnau (2017).
Once again the Bartlett test
(\ref{Bartlett_test})
will be applicable in the generalized log-normal model, which can be proved by following the proof of Harnau (2017).

In practice, we can fit seperate log-normal models to each group, that is
$y_{ij\ell}$
is assumed
$\mathsf{N}(\mu_{ij\ell},\omega_\ell^2)$.
If the Bartlett test does not reject the hypothesis of common variance we then arrive at a model where
$y_{ij\ell}$
is assumed
$\mathsf{N}(\mu_{ij\ell},\omega^2)$.
This model can be estimated by a single regression where the design matrix is block diagonal,
$X^m=\mathrm{diag}(X_1,X_2,\dots ,X_m)$
of dimension $p_\cdot=\sum_{\ell=1}^m p_\ell$.
We then compare the models with design matrices $X^m$ and the original $X$
of the maintained model through an $\mathsf{F}$-test.

%%%%%%%%%%%%%%%%%%%%%%%%%%%%%%%%%%%%%%%%%%%%%
%%%%%%%%%%%%%%%%%%%%%%%%%%%%%%%%%%%%%%%%%%%%%
%%%%%%%%%%%%%%%%%%%%%%%%%%%%%%%%%%%%%%%%%%%%%
\section{Empirical illustration}
\label{s:emp}

We apply the theory to the insurance run-off triangle shown in
Table \ref{tab:XLdata}.
All R (2017) code is given in the supplementary material.
We use the R packages \texttt{apc}, see
Nielsen (2015) and \texttt{ChainLadder}, see Gesmann et. al. (2015).
First, we apply the proposed inference and estimation procedures to the data.
This is followed first by distribution forecast and then by an analysis of the model specification.

%%%%%%%%%%%%%%%%%%%%%%%%%%%%%%%%%%%%%%%%%%%%%
\subsection{Inference and estimation}

We apply the log-normal model to the data and consider
three nested parametrizations:
\begin{itemize*}
    \item[apc]
    age-period-cohort model = extended chain-ladder
    \item[ac]                                                          
    age-cohort model = chain-ladder
    \item[ad]
    age-drift model = chain-ladder with a linear accident year effect        
\end{itemize*}
Table \ref{tab:XL:anova} shows an analysis of variance.
This conforms with the exact distribution theory in
\S\ref{ss:logN:stats}
and the asymptotic distribution theory in
Theorems
\ref{t:asymp_est},
\ref{t:asymp:inf}
in
\S\ref{ss:logN:asymp:inf}.

\begin{table}[t]
\centering
\begin{tabular}{lcccccc}
                                                                    \hline
    $sub$   & $-2\log\mathsf{L}$
                        & $df_{sub}$
                                & $\mathrm{F}_{sub,apc}$
                                        & $\mathsf{p}$
                                                & $\mathrm{F}_{sub,ac}$
                                                        & $\mathsf{p}$
                                                                \\  \hline
    apc     & 170.003   & 153   &       &       &       &       \\
    ac      & 179.873   & 171   & 0.41  & 0.984 &       &       \\
    ad      & 258.570   & 189   & 2.23  & 0.000 & 4.32  & 0.000 \\                                                                                                                                                                     \hline
\end{tabular}
\caption{Analysis of variance for the US casualty data}
\label{tab:XL:anova}
\end{table}

First, we test the chain-ladder model (ac for age-cohort)
against the extended chain-ladder model (apc for age-period-cohort)
with
$\mathsf{p}=0.984$.
The chain-ladder hypothesis is clearly not rejected at a conventional $5\%$ test level.
Next, we test the further restriction (ad for age-drift) that the row differences are
constant, that is
$\Delta^2\alpha_i=0$.
We get
$\mathsf{p}=0.000$
and
$\mathsf{p}=0.000$
when testing against the
apc and ac models respectively.
This suggests that a further reduction of the model is not supported.
In summary, the analysis of variance indicates that it is adequate to proceed with a
chain-ladder specification
and thereby ignore calendar effects. 

\begin{table}[ht]
\centering
\begin{tabular}{lrrlrr}
                                                                                    \hline
                        & estimate
                                & $se_t$&                   & estimate
                                                                    & $se_t$    \\  \hline
    $\mu_{11}$          &  7.660&   0.138    &                   &       &           \\
    $\Delta\alpha_2$    &  0.289& 0.134      & $\Delta\beta_2$   &  2.272&   0.134        \\
    $\Delta\alpha_3$    &  0.163& 0.136      & $\Delta\beta_3$   &  0.933&  0.136         \\
    $\Delta\alpha_4$    & -0.265& 0.140      & $\Delta\beta_4$   &  0.236&  0.140         \\
    $\Delta\alpha_5$    &  0.150& 0.144      & $\Delta\beta_5$   &  0.089&  0.144        \\
    $\Delta\alpha_6$    & -0.374& 0.148      & $\Delta\beta_6$   & -0.176&  0.148         \\
    $\Delta\alpha_7$    & -0.199& 0.153      & $\Delta\beta_7$   & -0.144&  0.153         \\
    $\Delta\alpha_8$    & -0.009& 0.159      & $\Delta\beta_8$   & -0.428&  0.159         \\
    $\Delta\alpha_9$    & -0.005& 0.165      & $\Delta\beta_9$   & -0.301&  0.165         \\
    $\Delta\alpha_{10}$ & -0.132& 0.172      & $\Delta\beta_{10}$& -0.400&  0.172         \\
    $\Delta\alpha_{11}$ & -0.022& 0.180      & $\Delta\beta_{11}$& -0.190&  0.180         \\
    $\Delta\alpha_{12}$ & -0.473& 0.190      & $\Delta\beta_{12}$& -0.242&  0.190         \\
    $\Delta\alpha_{13}$ & -0.438& 0.200      & $\Delta\beta_{13}$& -0.260&  0.200         \\
    $\Delta\alpha_{14}$ &  0.296& 0.214      & $\Delta\beta_{14}$& -0.555&  0.214         \\
    $\Delta\alpha_{15}$ &  0.311& 0.230      & $\Delta\beta_{15}$& -0.303&  0.230         \\
    $\Delta\alpha_{16}$ & -0.269& 0.250      & $\Delta\beta_{16}$&  0.406&  0.250         \\
    $\Delta\alpha_{17}$ &  0.142& 0.277      & $\Delta\beta_{17}$& -0.895&  0.277         \\
    $\Delta\alpha_{18}$ &  0.202& 0.316      & $\Delta\beta_{18}$&  0.117&  0.316         \\
    $\Delta\alpha_{19}$ & -0.093& 0.378      & $\Delta\beta_{19}$& -0.383& 0.378          \\
    $\Delta\alpha_{20}$ &  0.873& 0.508      & $\Delta\beta_{20}$& -0.273& 0.508          \\
    $s^2$               &  0.169&       & $RSS$             & 28.956&           \\  \hline    
\end{tabular}
\caption{Estimates for the US casualty data for the log-normal chain-ladder (ac).}
\label{tab:XL:LNestimates}
\end{table}

Table \ref{tab:XL:LNestimates} shows the estimated parameters for
the log-normal model with chain-ladder structure (ac).
We report standard errors
$se_t$
following
Theorem
\ref{t:asymp:inf:t}. They are the same for $\Delta\alpha$ and $\Delta\beta$ due to symetry of $(X'X)^{-1}$ at the diagonal.
These follow a $\mathsf{t}$-distribution with
$n-p=171$ degrees of freedom,
since the triangle has dimension $k=20$ and
$n=k(k+1)/2=210$
and
$p=2k-1=39$.
The corresponding two-sided 95\% critical values are 1.97.
We also report the degrees of freedom corrected estimate, $s^2$, for $\omega^2$.
We see that many of the development year effects
$\Delta\beta$, in particular
$\Delta\beta_2$,
are significant. The first few development year effects are positive, which matches the
increases seen in first few columns of the data in
Table \ref{tab:XLdata}.
At the same time many the accident year effects
$\Delta\alpha$
are not individually significant, although they are jointly significant as seen in
Table \ref{tab:XL:anova}.
The signs of the
$\Delta\alpha$'s
match the relative increase or decrease of the amounts seen in the rows of 
Table \ref{tab:XLdata}.

In Appendix \ref{s:tables}
we present a further Table \ref{tab:XL:LNestimates:apc}
with estimates. These are the estimated parameters for the log-normal model
with an extended chain-ladder structure (apc)
as in
\S\ref{ss:logN:apc}.
These will be used for the simulation study.
The $\Delta^2\gamma$-coefficients measure the calendar effect and are restricted to zero in the
chain-ladder model.

%%%%%%%%%%%%%%%%%%%%%%%%%%%%%%%%%%%%%%%%%%%%%
\subsection{Distribution forecasting}

\begin{table}[t]
\centering
\begin{tabular}{rrrrrrrrrr}
                                                                                                                \hline
            & \multicolumn{3}{l}{generalized log-normal}            & \multicolumn{3}{l}{over-dispersed Poisson}           & \multicolumn{3}{l}{bootstrap}     \\ 
    \vspace{-12pt}                                                                                                      \\  
    $i$     & Reserve   & $\dfrac{se}{\text{Res}}$
                                    & $\dfrac{99.5\%}{\text{Res}}$
                                                & Reserve   & $\dfrac{se}{\text{Res}}$
                                                                        & $\dfrac{99.5\%}{\text{Res}}$
                                                                                    & Reserve   & $\dfrac{se}{\text{Res}}$
                                                                                                            & $\dfrac{99.5\%}{\text{Res}}$
                                                                                                                        \\
    \vspace{-10pt}                                                                                                      \\  \hline
     2      &   1871    & 0.55      & 2.43      & 1368      & 1.81      & 5.71      & 1345      & 1.99      & 9.93      \\
     3      &   5099    & 0.37      & 1.96      & 4476      & 0.92      & 3.40      & 4415      & 0.97      & 4.63      \\
     4      &   7171    & 0.30      & 1.77      & 6925      & 0.69      & 2.78      & 6830      & 0.71      & 3.56      \\
     5      &  11699    & 0.26      & 1.66      & 10975     & 0.54      & 2.41      & 10846     & 0.56      & 2.90      \\
     6      &  13717    & 0.24      & 1.64      & 14941     & 0.44      & 2.14      & 14767     & 0.45      & 2.50      \\
     7      &  14344    & 0.22      & 1.58      & 18337     & 0.39      & 2.01      & 18147     & 0.40      & 2.29      \\
     8      &  18377    & 0.21      & 1.54      & 24487     & 0.34      & 1.87      & 24233     & 0.35      & 2.09      \\
     9      &  25488    & 0.21      & 1.54      & 31876     & 0.29      & 1.76      & 31607    & 0.30      & 1.93     \\
     10     &  30525    & 0.20      & 1.53      & 35567     & 0.28      & 1.72      & 35270     & 0.28      & 1.87      \\
     11     &  40078    & 0.20      & 1.53      & 48595     & 0.24      & 1.63      & 48176     & 0.25      & 1.73      \\  
     12     &  32680    & 0.20      & 1.53      & 42027     & 0.26      & 1.68      & 41659     & 0.27      & 1.80      \\
     13     &  28509    & 0.21      & 1.54      & 37114     & 0.28      & 1.74      & 36814     & 0.29      & 1.88      \\
     14     &  51761    & 0.21      & 1.55      & 66977     & 0.22      & 1.58      & 66554     & 0.23      & 1.69      \\
     15     &  98748    & 0.22      & 1.58      & 102982    & 0.20      & 1.51      & 102282    & 0.20      & 1.59      \\
     16     & 100331    & 0.23      & 1.60      & 136647    & 0.19      & 1.51      & 135880    & 0.20      & 1.59      \\
     17     & 149813    & 0.24      & 1.64      & 164318    & 0.22      & 1.56      & 163500    & 0.22      & 1.68      \\
     18     & 221550    & 0.26      & 1.69      & 218874    & 0.25      & 1.66      & 218115    & 0.26      & 1.83      \\
     19     & 229481    & 0.30      & 1.79      & 166120    & 0.49      & 2.29      & 166431    & 0.51      & 2.84      \\
     20     & 575343    & 0.41      & 2.06      & 337001    & 0.94      & 3.46      & 353628    & 1.03      & 4.91      \\
     total  & 1656586   & 0.16      & 1.42      & 1469605   & 0.23      & 1.60      & 1480500   & 0.26      & 1.95      \\  \hline
\end{tabular}
\caption{Forecasting for the US casualty data using the generalized log-normal, the over-dispersed Poisson model and the
bootstrap.  The bootstrap simulation is based on $10^5$ repetitions.}
\label{tab:XL:LNforecasts}
\end{table}

Table \ref{tab:XL:LNforecasts}
shows forecasts of reserves for the US casualty data
in different accident years, i.e.\ the row sums in the lower triangle
$\mathcal{J}$.
We report results from the generalized log-normal chain-ladder model (GLN),
the over-dispersed Poisson chain-ladder (ODP) and
England (2002) bootstrap (BS).
For each method, we present a point forecast of the reserve,
the standard error over point forecast ($se$/Res)
and the 1 in 200 over point forecast values (99.5\%/Res).

For the generalized log-normal chain-ladder model we use the asymptotic distribution forecast in
(\ref{LN_distribution_forecast}).
For the over-dispersed Poisson model we use the asymptotic distribution forecasts
from
Harnau and Nielsen (2017, equation 11).
For the bootstrap
we use 
the ChainLadder package by Gesmann et al (2005), based on the method described in England (2002).
We apply $10^5$ bootstrap draws using the gamma option.

Table \ref{tab:XL:LNforecasts}
shows that the
over-dispersed Poisson forecasts are similar to the bootstrap.
Their point forecasts are smaller than that of the generalized log-normal model.
This is in part due to the additional factor $\exp(s^2/2)=\exp(0.169/2)=1.088$
in the generalized log-normal point forecast.
The difference seems large compared to the authors' experience with other data.
It is possibly due to the relatively large dimension of the triangle, so that there are more
degrees of freedom to pick up differences
between the over-dispersed Poisson and the generalized log-normal models.

The standard error and 99.5\% quantiles over reserve ratios are generally lower
and less variable for the generalized log-normal chain-ladder model. This is especially pronounced
for early accident years and the latest accident year.

\setlength{\tabcolsep}{-7pt}
\begin{figure}[t]
\centering
\begin{tabular}{ccc}
      \includegraphics[width=56.5mm]{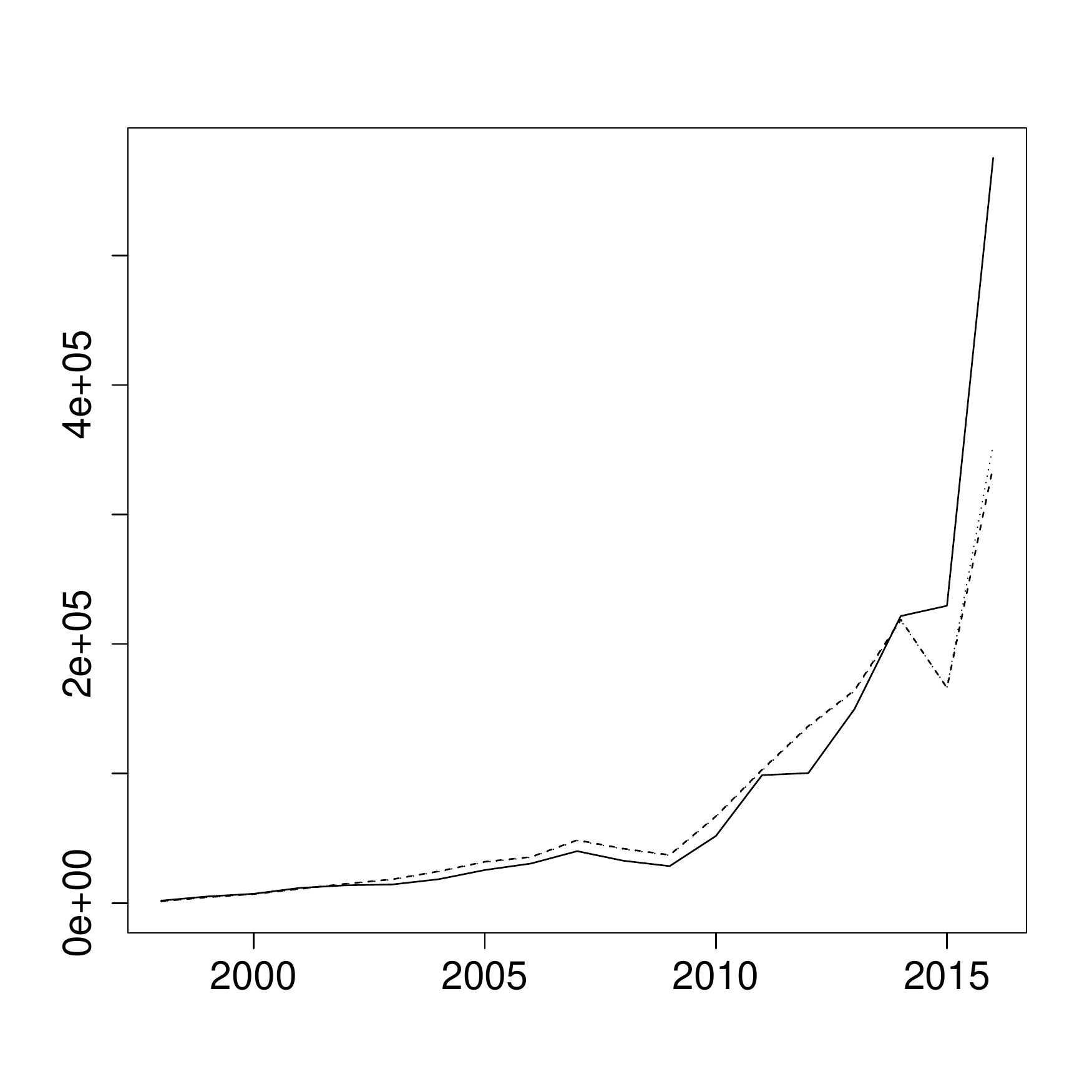}    &
      \includegraphics[width=56.5mm]{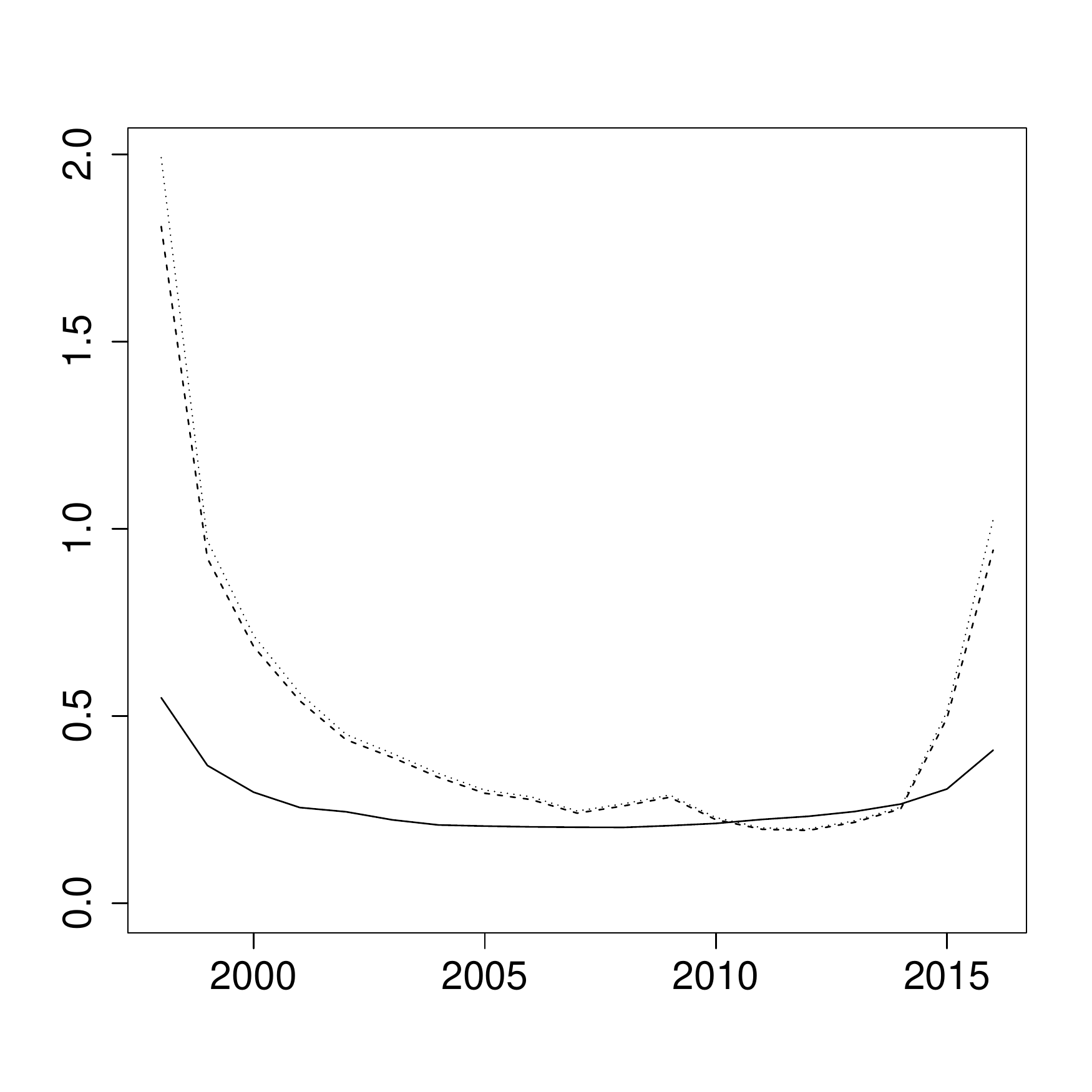} &
      \includegraphics[width=56.5mm]{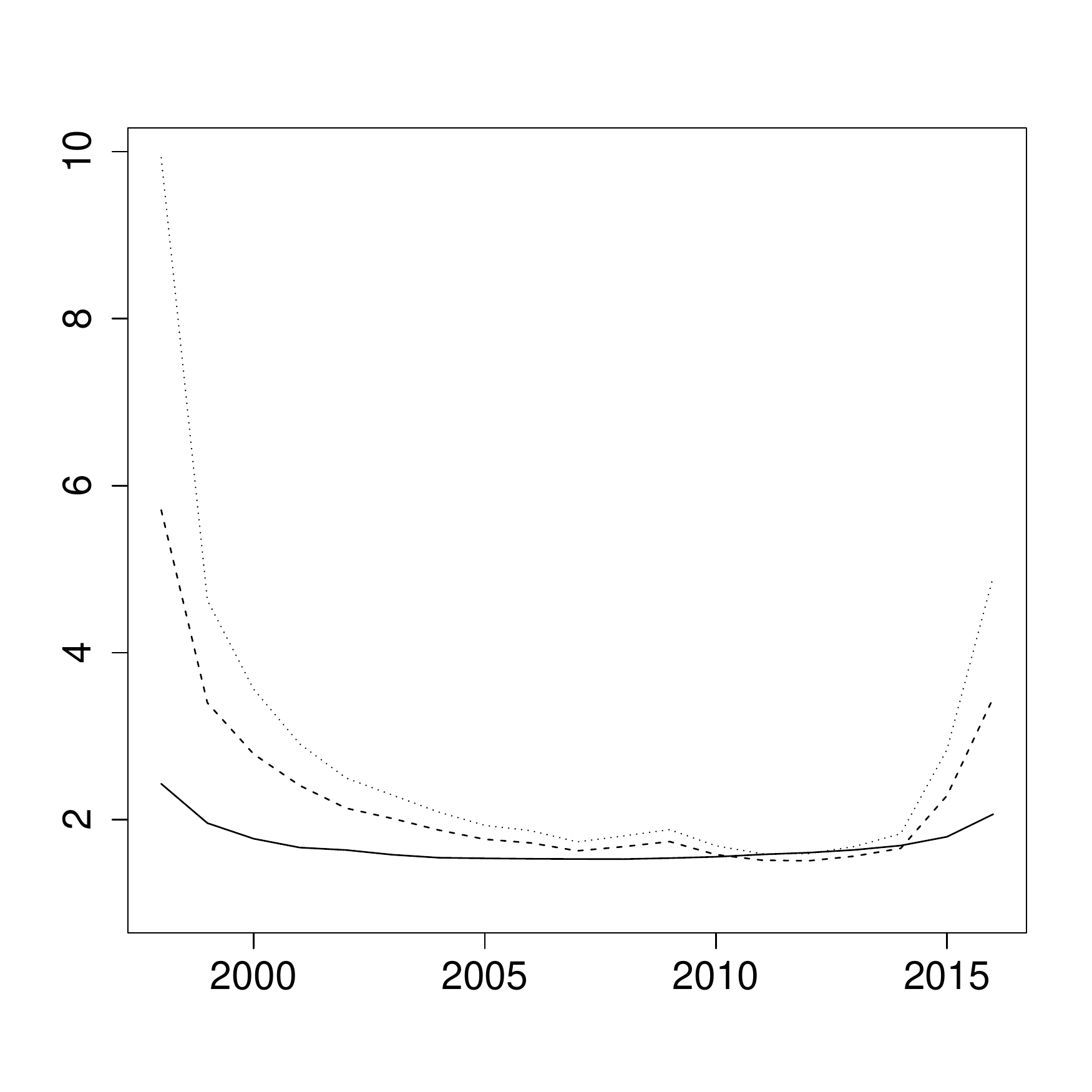}
      \vspace{-10pt}
                                                        \\
      \vspace{-5pt}
      \small    (a) Reserve        &
      \small    (b) se/Reserve     &
      \small    (c) 99.5\%/Reserve \\
\end{tabular}
\caption{Illustration of the forecasts in Table \protect{\ref{tab:XL:LNforecasts}} for the US casualty data.
    Solid line is the generalized log-normal forecast.
    Dashed line is the over-dispersed Poisson forecast.
    Dotted line is the bootstrap forecast.
    Panel (a) shows the reserves against accident year $i$.
    Panel (b) shows the standard error to reserve ratio.
    Panel (c) shows the 99.5\% quantile to reserve ratio.}
\label{fig:XL:LNestimates}
\end{figure}
\setlength{\tabcolsep}{6pt}

Figure \ref{fig:XL:LNestimates} shows the trends of the reserve and
standard error and 99.5\% quantile over reserve ratios for the three methods.
The point forecast trends are similar for models, showing an increasing trend with accident year
as expected.
The ratios are seen to be flatter for the generalized log-normal model.
This is related to the assumption of the
generalized log-normal chain-ladder model
that standard deviation to mean ratio is constant across the entries, while
the variance to mean ratio is assumed constant for the over-dispersed Poisson model and the bootstrap.

%%%%%%%%%%%%%%%%%%%%%%%%%%%%%%%%%%%%%%%%%%%%%

\begin{table}[t]
\centering
\begin{tabular}{rrr|rrr|rrr}
\multicolumn{3}{l}{Full triangle}                   &   \multicolumn{3}{l}{Leave 1 out}                 &   \multicolumn{3}{l}{Leave 2 out}                 \\
\multicolumn{9}{l}{generalized log-normal}                                                                  \\
$i$ &   $\dfrac{se}{Res}$    &   $\dfrac{99.5\%}{Res}$    &
$i$ &   $\dfrac{se}{Res}$    &   $\dfrac{99.5\%}{Res}$    &
$i$ &   $\dfrac{se}{Res}$    &   $\dfrac{99.5\%}{Res}$    \\
16  &    0.23   &    1.60   &   15  &    0.23   &    1.61   &   14  &    0.23   &    1.61   \\
17  &    0.24   &    1.64   &   16  &    0.25   &    1.64   &   15  &    0.25   &    1.64   \\
18  &    0.26   &    1.69   &   17  &    0.27   &    1.69   &   16  &    0.27   &    1.69   \\
19  &    0.30   &    1.79   &   18  &    0.31   &    1.80   &   17  &    0.31   &    1.80   \\
20  &    0.41   &    2.06   &   19  &    0.41   &    2.07   &   18  &    0.41   &    2.07   \\
all &    0.16   &    1.42   &   all &    0.13   &    1.33   &   all &    0.12   &    1.31   \\
\hline                  &                       &                       \\
\multicolumn{9}{l}{over-dispersed Poisson}                                                                 \\
%$i$ &   SE / Res    &   99.5\% / Res    &   $i$ &   SE / Res    &   99.5\% / Res    &   $i$ &   SE / Res    &   99.5\% / Res    \\
16  &    0.19   &    1.51   &   15  &    0.20   &    1.53   &   14  &    0.22   &    1.58   \\
17  &    0.22   &    1.56   &   16  &    0.22   &    1.56   &   15  &    0.24   &    1.62   \\
18  &    0.25   &    1.66   &   17  &    0.28   &    1.74   &   16  &    0.28   &    1.72   \\
19  &    0.49   &    2.29   &   18  &    0.48   &    2.25   &   17  &    0.48   &    2.24   \\
20  &    0.94   &    3.46   &   19  &    1.38   &    4.61   &   18  &    1.51   &    4.94   \\
all &    0.23   &    1.60   &   all &    0.20   &    1.53   &   all &    0.20   &    1.52   \\
\hline                  &                       &                       \\
\multicolumn{9}{l}{bootstrap}                                                                  \\
%$i$ &   SE / Res    &   99.5\% / Res    &   $i$ &   SE / Res    &   99.5\% / Res    &   $i$ &   SE / Res    &   99.5\% / Res    \\
16  &    0.20   &    1.59   &   15  &    0.21   &    1.62   &   14  &    0.23   &    1.70   \\
17  &    0.22   &    1.68   &   16  &    0.22   &    1.68   &   15  &    0.24   &    1.75   \\
18  &    0.26   &    1.83   &   17  &    0.29   &    1.97   &   16  &    0.28   &    1.92   \\
19  &    0.51   &    2.84   &   18  &    0.49   &    2.78   &   17  &    0.49   &    2.77   \\
20  &    1.03   &    4.91   &   19  &    1.49   &    6.69   &   18  &    1.66   &    7.45   \\
all &    0.26   &    1.95   &   all &    0.23   &    1.81   &   all &    0.22   &    1.79   \\
\hline                                                                      
\end{tabular}
\caption{Recursive forecasting for the US casualty data in the latest 5 accident years. 
The bootstrap simulation is based on $10^5$ repetitions.}
\label{tab:XL:LNforecastsStability}
\end{table}

%%%%%%%%%%%%%%%%%%%%%%%%%%%%%%%%%%%%%%%%%%%%%
\subsection{Recursive distribution forecasting}

To check the robustness of the model we apply the distribution forecasting recursively.
Thus, we apply the distribution forecast to subsets of the triangle.

In this way,
Table \ref{tab:XL:LNforecastsStability}
shows standard error and 99.5\% over reserve ratios.
It has 9 panels,
where the rows are for the asymptotic generalized log-normal model,
the over-dispersed Poisson model
and the bootstrap, respectively.
In the first column we show the ratios for the last 5 accident years based on the full triangle.
These numbers are the same as those in
Table \ref{tab:XL:LNforecasts}.
In the second column we omit the last diagonal of the data triangle to get a $k-1=19$ dimensional triangle.
We then forecast the last 5 accident years relative to that triangle.
In the third column we omit the last two diagonals of the data triangle to get a $k-2=18$ dimensional triangle.

We see that the generalized log-normal forecasts are stable for all years.
The over-dispersed Poisson and bootstrap forecasts
are less stable in the latest accident year.
This is possibly because of instability in the corners of the
data triangle shown in
Table \ref{tab:XLdata},
that may be dampened when taking logs.
Alternatively, it could be attributed to a better fit of the log-normal model across the entire triangle.
We will explore the model specification using formal tests in the next section.

%%%%%%%%%%%%%%%%%%%%%%%%%%%%%%%%%%%%%%%%%%%%%
\subsection{Model selection}

\begin{table}[t]
\centering
\begin{tabular}{ccccccccc}
    \hline  
    &   \multicolumn{4}{l}{generalized log-normal}
    &   \multicolumn{4}{l}{over-dispersed Poisson}
    \\
    splits
    & $LR^\omega/C$ & $\mathsf{p}$ & ${F}$ & $\mathsf{p}$ 
    & $LR^\omega/C$ & $\mathsf{p}$ & ${F}$ & $\mathsf{p}$ 
    \\
    \hline
    (a) & 6.29 & 0.012 & 1.34 & 0.030 & 11.68 & 0.001 & 2.29 & 0.000 \\
    (b) & 4.70 & 0.095 & 1.55 & 0.005& 11.63 & 0.003 & 4.17 & 0.000 \\
    (c) & 1.12 & 0.291 & 1.33 & 0.037 & 15.07 & 0.000 & 2.30 & 0.000 \\
    \hline
\end{tabular}
\caption{Bartlett tests for common dispersion and F tests for common mean parameters.}
\label{tab:Bartlett}
\end{table}

We now apply the specification test outlined in
\S\ref{ss:logN:asymp:spec}
for the log-normal model and in
Harnau (2017)
for the over-dispersed Poisson model.
For the tests we split the data triangle of
Table \ref{tab:XLdata}
as outlined in
Figure \ref{F:Test_Triangles_cut}:
\begin{itemize*}
    \item[(a)]
        a horizontal split with the first 6 rows in one group and the last 14 rows in a second group.
    \item[(b)]
        a horizontal and diagonal split with
        the first 10 diagonals in one group, the last 10 rows in a second group and the remaining entries in a third group.
    \item[(c)]
        a diagonal split with the first 14 diagonals in one group and the last 6 diagonals in a second group.
\end{itemize*}
For each split we estimate a chain-ladder structure separately for each sub-group. We
then compute the Bartlett test statistic
$LR^\omega/C$
from
(\ref{Bartlett})
for a common variance across groups.
Given a common variance we also compute an $F$-statistic
for common chain-ladder structure in the mean.

For each of the generalized log-normal and over-dispersed Poisson model we are conducting 6 tests.
When chosing the size of each individual test, that is the probability of falsely rejecting the hypothesis,
we would have to keep in mind the overall size of rejecting any of the hypotheses.
If the test statistics were independent and the individual tests were conducted at level $p$ the overall size would be
$1-(1-p)^6\approx 6p$ by binomial expansion, see also
Hendry and Nielsen (2007, \S 9.5).
Thus, if the individual tests are conducted at a 1\% level
we would expect the overall size to be about 5\%.  
At present we have no theory for a more formal calculation of the joint size of the tests.

Starting with the log-normal model we see that there is only moderate evidence against model.
The worst cases are that variance differs across the (a) split and the chain-ladder structure
differs across the (b) split.  In contrast, the over-dispersed Poisson model
is rejected by all 6 tests.

%%%%%%%%%%%%%%%%%%%%%%%%%%%%%%%%%%%%%%%%%%%%%
%%%%%%%%%%%%%%%%%%%%%%%%%%%%%%%%%%%%%%%%%%%%%
%%%%%%%%%%%%%%%%%%%%%%%%%%%%%%%%%%%%%%%%%%%%%
\section{Simulation}
\label{s:sim}

In
Theorems
\ref{t:asymp:inf}
and
\ref{t:asymp_forecast}
we presented asymptotic results for
inference and distribution forecasting.
We now apply simulation to investigate the quality of these asymptotic approximations.

%%%%%%%%%%%%%%%%%%%%%%%%%%%%%%%%%%%%%%%%%%%%%
\subsection{Test statistic}\label{s:TestStatistics}

We assess the finite sample performance of the $F$-tests proposed in
Theorem \ref{t:asymp:inf}
and applied in
Table \ref{tab:XL:anova}.
We simulate under the null hypothesis of a chain-ladder specification, ac,
as well as under the alternative hypothesis of an extended chain-ladder specification, apc.
We choose the distribution to be log-normal so, to be specific, we actually illustrate the
well-known exact distribution theory for regression analysis.
Theorem \ref{t:asymp:inf}
also applies
for infinitely divisible distributions that are not log-normal but satisfy
Assumptions
\ref{as:HN:inf_div}
and
\ref{as:HN:inf_div:LN}.
Such infinitely divisible distributions are, however, not easily generated.
The real point of the simulations is therefore to illustrate the small variance asymptotics
in
Theorem \ref{t:asymp:inf}
by showing that power increases with shrinking variance.

The data generating processes are constructed from the US casualty data as follows.
We consider a $k=20$ dimensional triangle.
We assume that the variables
$Y_{ij}$
in the upper triangle
$\mathcal{I}$
are independent log-normal distributed, so that
$y_{ij}=\log(Y_{ij})$
is normal with mean
$\mu_{ij}$
and variance
$\sigma^2$.
Under the null hypothesis of a chain-ladder specification,
$\mathsf{H}_{ac}$,
then
$\mu_{ij}$
is defined from
(\ref{muxi})
where the parameters 
$\mu_{ij}$
are chosen to match those of
Table \ref{tab:XL:LNestimates}.
We also choose
$\sigma^2$
to match the estimate
$s^2$
from 
Table \ref{tab:XL:LNestimates},
but multiplied by a factor
$v^2$
where $v$ is chosen as
$2, 1, 1/2$
to capture the small-variance asymptotics.
Under the alternative, we apply the extended chain-ladder specification
$\mathsf{H}_{apc}$
where the parameters are chosen to match those of
Table \ref{tab:XL:LNestimates:apc}.
In all cases we
draw $10^5$ repetitions.

\begin{table}[t]
\centering
\begin{tabular}{lrrrrrr}                                                                \hline

            & \multicolumn{3}{l}{Size under $\mathsf{H}_{ac}$}
                                                & \multicolumn{3}{l}{Power under $\mathsf{H}_{apc}$}
                                                                                    \\ 
    Confidence level    & 1.00\%    & 5.00\%    & 10.00\%   & 1.00\%   & 5.00\%   & 10.00\%   \\  \hline
    $v=2$   & 1.01\%    & 5.00\%    &  10.16\%   & 2.26\%   & 9.03\%   & 16.31\%   \\
    $v=1$   & 0.98\%    & 5.07\%    &  10.07\%   & 10.49\%   & 27.51\%   & 40.22\%   \\ 
    $v=0.5$ & 0.99\%    & 5.09\%    &  10.05\%   & 78.03\%   & 92.17\%   & 96.07\%   \\  \hline 
\end{tabular}
\caption{Simulated performance of $\textnormal{F}$ test based on $10^5$ draws.
The Monte Carlo standard error less than 0.01.}
\label{tab:sim:test}
\end{table}

We note that the $\mathsf{F}(18,153)$-distribution is exact under the null hypothesis,
since we are operating on the log-scale and simulate normal variables so that
standard regression theory applies.
Indeed,
Table \ref{tab:sim:test}
shows that simulated size (type I error) is correct apart from Monte Carlo standard error.
We check this for at the 1\%, 5\% and 10\% level for
$v=2, 1, 1/2$.

Under the alternative we simulate power (unity minus type II error).
The exact distribution is a non-central $\mathsf{F}$-distribution.
The simulations show that the power increases for shrinking variance $v^2\omega^2$ and for increasing
level (type I error) of the test.

We can also illustrate
the increasing power with shrinking variance through the following analytic example.
Suppose we consider variables
$Z_1,\dots ,Z_n$
that are independent
$\mathsf{N}(\mu,\omega^2)$-distributed.
Then the parameters are estimated by
$\hat\mu=\bar{Z}$
and
$s^2=(n-1)^{-1}\sum_{i=1}^n (Z_i - \bar{Z})^2$.
The $\mathsf{t}$-statistic for $\mu=0$ has the expansion
$$
    \frac{\hat\mu -  0}{\sqrt{s^2/(n-1)}}
    =
    \frac{\hat\mu -\mu}{\sqrt{s^2/(n-1)}}
    +
    \frac{    \mu -  0}{\sqrt{s^2/(n-1)}}
    .
$$
The first term is $\mathsf{t}$ distributed with $(n-1)$ degrees of freedom
regardless of the value of $\mu$.
The second term is zero under the hypothesis $\mu=0$.
Under the alternative $\mu\ne 0$ the second term is non-zero and measures non-centrality
so that the overall $\mathsf{t}$-statistic is non-central
$\mathsf{t}$.
In standard asymptotic theory $n$ is large so that for fixed $\mu$, $\omega$ then
$s^2$ is consistent for $\omega^2$ and the
second term is close to
$\mu/\surd{\omega^2 / (n-1)}=(\mu/\omega)\surd (n-1)$.
Due to the $(n-1)$-factor 
the non-centrality diverges, so that the power increases to unity and the test is consistent.
In the small variance asymptotics $\omega^2$ shrinks to zero while $n$ is fixed.
Then
$s^2$ vanishes, see
Theorem \ref{t:asymp:inf},
and the non-centrality diverges in a similar way even though $n$ is fixed.

%%%%%%%%%%%%%%%%%%%%%%%%%%%%%%%%%%%%%%%%%%%%%
\subsection{Forecasting}

We assess the finite sample performance of the asymptotic distribution forecasts proposed in
Theorem \ref{t:asymp_forecast}
and applied in
Table \ref{tab:XL:LNforecasts}.
These asymptotic distribution forecasts are compared to
the over-dispersed Poisson forecast of
Harnau and Nielsen (2017)
and the bootstrap of
England and Verrall (1999)
and
England (2002).
Two different log-normal chain-ladder data generating processes are used.
First, we apply the estimates from the US casualty data
so that the parameters 
are chosen to match those of
Table \ref{tab:XL:LNestimates}.
As before the variance $\omega^2$ is multiplied by a factor $v^2$ where
$v=2,1,1/2$.
We have seen that the over-dispersed Poisson model is poor for this data set and
we will expect the generalized log-normal distribution forecasts to be superior.
Secondly, we obtain similar estimates for the
Taylor and Ashe (1983)
data, see also
Harnau and Nielsen (2017, Table 1).
For those data the generalized log-normal model and the over-dispersed Poisson model
provide equally good fits so that the different distributions forecasts should be more similar
in performance.

We first compare the asymptotic distribution forecast from
Theorem \ref{t:asymp_forecast}
with the exact forecast distribution.
This is done by simulating log-normal chain-ladder for both the upper and the lower
triangles,
$\mathcal{I}$
and
$\mathcal{J}$.
The true forecast error distribution is then based on
$Y_\mathcal{A}-\tilde{Y}_\mathcal{A}$,
where
$Y_\mathcal{A}$
is computed from the simulated lower triangle
$\mathcal{J}$
while
$\tilde{Y}_\mathcal{A}$
is the log-normal point forecast computed from the upper triangle data 
$\mathcal{I}$.
We compute the true forecast error
$Y_\mathcal{A}-\tilde{Y}_\mathcal{A}$
for each simulation draw and report mean, standard error and
quantiles of the draws.
This is done for the entire reserve, so that $\mathcal{A}=\mathcal{J}$.
The asymptotic theory in
Theorem \ref{t:asymp_forecast}
provides a $\mathsf{t}$-approximation, so that for each draw of the upper triangle
$\mathcal{I}$, we also compute
mean, standard error and quantiles
from the $\mathsf{t}$-approximations and report averages over the draws.

The first panel of 
Table \ref{tab:sim:forecast}
compares the simulated actual forecast distribution,
$true^{GLN}$,
with the simulated $\mathsf{t}$-approximations,
$t^{GLN}$.
We see that with shrinking variance factor $v$ then the overall
forecast distribution becomes less variable and the
$\mathsf{t}$-approximation becomes relatively better.
The 
$\mathsf{t}$-approximation
is symmetric and does not fully capture
the asymmetry of the actual distribution.   We note that the performance of
the
$\mathsf{t}$-approximation
is better in the upper tail than the lower tail, which is beneficial when we are
interested in 99.5\% value at risk.

\begin{table}
\centering
\begin{tabular}{llrrrrrrrrr}
                                                                                        \hline
    &       &   \multicolumn{2}{l}{Moments} &                   \multicolumn{7}{l}{Quantiles}                                           \\  
v   &       &   Mean    &   SE  &   0.5\%   &   1\% &   5\% &   50\%    &   95\%    &   99\%    &   99.5\%  \\  \hline
    &   \multicolumn{10}{l}{generalized log-normal (GLN)}                                                                             \\  
2   &   $true^{GLN}$ &   3.0 &   12.6    &   -55.1   &   -42.6   &   -18.5   &   5.4 &   17.2    &   22.2    &   24.4    \\  
    &   $t^{GLN}$    &   0.0 &   7.9 &   -20.7   &   -18.7   &   -13.1   &   0.0 &   13.1    &   18.7    &   20.7    \\  
1   &   $true^{GLN}$ &   0.5 &   3.3 &   -11.2   &   -9.5    &   -5.5    &   0.9 &   5.0 &   6.5 &   7.0 \\  
    &   $t^{GLN}$    &   0.0 &   3.0 &   -7.7    &   -6.9    &   -4.9    &   0.0 &   4.9 &   6.9 &   7.7 \\  
0.5 &   $true^{GLN}$ &   0.1 &   1.4 &   -4.1    &   -3.6    &   -2.3    &   0.2 &   2.3 &   3.0 &   3.3 \\  
    &   $t^{GLN}$    &   0.0 &   1.4 &   -3.6    &   -3.2    &   -2.3    &   0.0 &   2.3 &   3.2 &   3.6 \\  \hline
v   &   \multicolumn{10}{l}{over-dispersed Poisson (ODP) and bootstrap (BS)}                                                                             \\  
2   &   $true^{ODP}$    &   7.7 &   10.5    &   -37.9   &   -28.5   &   -10.0   &   9.3 &   20.3    &   25.4    &   27.3    \\  
    &   $t^{ODP}$   &   0.0 &   19.8    &   -51.6   &   -46.5   &   -32.8   &   0.0 &   32.8    &   46.5    &   51.6    \\  
    &   BS  &   -15.4   &   2631.6  &   -683.1  &   -350.8  &   -78.9   &   3.3 &   55.8    &   313.3   &   643.1   \\  
1   &   $true^{ODP}$    &   1.3 &   3.2 &   -9.9    &   -8.3    &   -4.5    &   1.7 &   5.8 &   7.3 &   7.8 \\  
    &   $t^{ODP}$   &   0.0 &   7.9 &   -20.7   &   -18.6   &   -13.1   &   0.0 &   13.1    &   18.6    &   20.7    \\  
    &   BS  &   -1.8    &   123.4   &   -73.9   &   -50.1   &   -21.2   &   0.5 &   12.5    &   23.4    &   35.1    \\  
0.5 &   $true^{ODP}$    &   0.3 &   1.4 &   -4.0    &   -3.5    &   -2.2    &   0.4 &   2.5 &   3.3 &   3.6 \\  
    &   $t^{ODP}$   &   0.0 &   3.8 &   -9.8    &   -8.8    &   -6.2    &   0.0 &   6.2 &   8.8 &   9.8 \\  
    &   BS  &   -0.2    &   4.2 &   -15.4   &   -13.1   &   -7.5    &   0.1 &   5.9 &   9.1 &   10.3    \\  \hline
v   &   \multicolumn{10}{l}{root-mean-square-errors (rms)}                                                                            \\  
2   &   $rms^{GLN}$    &   3.0 &   8.3 &   38.7    &   28.8    &   12.5    &   5.4 &   11.9    &   16.3    &   18.1    \\  
    &   $rms^{ODP}$   &   7.7 &   13.8    &   29.7    &   29.9    &   28.2    &   9.3 &   20.9    &   31.8    &   35.9    \\  
    &   $rms^{BS}$  &   4284.4  &   135397.1    &   925.7   &   431.1   &   86.4    &   6.8 &   17.3    &   52.7    &   397.7   \\  
1   &   $rms^{GLN}$    &   0.5 &   1.1 &   4.5 &   3.6 &   1.9 &   0.9 &   1.8 &   2.6 &   2.9 \\  
    &   $rms^{ODP}$   &   1.3 &   5.1 &   11.9    &   11.3    &   9.2 &   1.7 &   8.0 &   12.2    &   13.8    \\  
    &   $rms^{BS}$  &   67.6    &   2132.3  &   79.5    &   48.4    &   18.2    &   1.2 &   5.4 &   6.1 &   18.8    \\  
0.5 &   $rms^{GLN}$    &   0.1 &   0.3 &   0.8 &   0.7 &   0.4 &   0.2 &   0.4 &   0.6 &   0.7 \\  
    &   $rms^{ODP}$   &   0.3 &   2.4 &   5.9 &   5.5 &   4.1 &   0.4 &   3.8 &   5.7 &   6.4 \\  
    &   $rms^{BS}$  &   0.6 &   3.0 &   11.9    &   10.0    &   5.5 &   0.3 &   2.4 &   2.7 &   5.7 \\  \hline
\end{tabular}
\caption{Simulation performance of distribution forecasts for the US casualty data.
Results in USD. The study is based on $10^5$ repetitions, and for each simulated upper triangle, the bootstrap is based on 999 simulations.}
\label{tab:sim:forecast}
\end{table}

\begin{table}
\centering
\begin{tabular}{llrrrrrrrrr}
    &       &   \multicolumn{2}{l}{Moments} &                   \multicolumn{7}{l}{Quantiles}                                           \\  
v        &   &   Mean    &   SE  &   0.5\%   &   1\% &   5\% &   50\%    &   95\%    &   99\%    &   99.5\%  \\  \hline
    &   \multicolumn{10}{l}{generalized log-normal (GLN)}                                                                             \\  
2   &   $true^{GLN}$ &   7.2 &   99.8    &   -372.9  &   -310.0  &   -170.0  &   20.4    &   140.6   &   187.5   &   206.2   \\  
    &   $t^{GLN}$    &   0.0 &   75.7    &   -205.7  &   -184.2  &   -127.7  &   0.0 &   127.7   &   184.2   &   205.7   \\  
1   &   $true^{GLN}$ &   1.7 &   31.8    &   -96.4   &   -83.7   &   -54.0   &   3.9 &   49.6    &   66.8    &   72.8    \\  
    &   $t^{GLN}$    &   0.0 &   29.7    &   -80.7   &   -72.2   &   -50.1   &   0.0 &   50.1    &   72.2    &   80.7    \\  
0.5 &   $true^{GLN}$ &   0.4 &   14.3    &   -39.6   &   -35.4   &   -23.9   &   0.9 &   23.0    &   31.7    &   34.4    \\  
    &   $t^{GLN}$    &   0.0 &   14.0    &   -38.0   &   -34.0   &   -23.6   &   0.0 &   23.6    &   34.0    &   38.0    \\  \hline
v   &   \multicolumn{10}{l}{over-dispersed Poisson (ODP) and bootstrap (BS)}                                                                             \\  
2   &   $true^{ODP}$    &   45.1    &   91.4    &   -297.9  &   -242.1  &   -116.8  &   56.9    &   168.2   &   213.5   &   230.8   \\  
    &   $t^{ODP}$   &   0.0 &   76.6    &   -208.4  &   -186.6  &   -129.4  &   0.0 &   129.4   &   186.6   &   208.4   \\  
    &   BS  &   -14.1   &   340.9   &   -414.3  &   -335.9  &   -193.8  &   -0.3    &   114.3   &   155.6   &   177.4   \\  
1   &   $true^{ODP}$    &   9.1 &   31.9    &   -89.8   &   -76.9   &   -46.8   &   11.4    &   56.9    &   73.5    &   79.6    \\  
    &   $t^{ODP}$   &   0.0 &   31.7    &   -86.1   &   -77.1   &   -53.5   &   0.0 &   53.5    &   77.1    &   86.1    \\  
    &   BS  &   -2.5    &   35.4    &   -109.5  &   -97.2   &   -64.6   &   0.1 &   50.5    &   68.2    &   74.1    \\  
0.5 &   $true^{ODP}$    &   2.1 &   14.7    &   -39.3   &   -34.7   &   -22.8   &   2.7 &   25.2    &   33.8    &   36.9    \\  
    &   $t^{ODP}$   &   0.0 &   15.1    &   -41.2   &   -36.9   &   -25.6   &   0.0 &   25.6    &   36.9    &   41.2    \\  
    &   BS  &   -0.6    &   16.5    &   -46.3   &   -41.6   &   -28.6   &   0.0 &   25.3    &   34.9    &   38.2    \\  \hline
v   &   \multicolumn{10}{l}{root-mean-square-errors (rms)}                                                                            \\  
2   &   $rms^{GLN}$    &   7.2 &   45.3    &   197.1   &   156.7   &   77.4    &   20.4    &   66.0    &   93.4    &   104.3   \\  
    &   $rms^{ODP}$   &   45.1    &   32.2    &   118.5   &   89.1    &   49.9    &   56.9    &   61.9    &   74.7    &   80.9    \\  
    &  $ rms^{BS}$  &   645.6   &   20322.2 &   415.1   &   259.0   &   126.9   &   57.4    &   168.6   &   107.6   &   107.7   \\  
1   &   $rms^{GLN}$    &   1.7 &   7.4 &   24.8    &   20.7    &   12.6    &   3.9 &   12.0    &   18.1    &   20.8    \\  
    &   $rms^{ODP}$   &   9.1 &   6.4 &   17.9    &   15.6    &   12.7    &   11.4    &   11.4    &   16.0    &   18.7    \\  
    &   $rms^{BS}$  &   11.7    &   8.6 &   36.0    &   32.7    &   23.7    &   11.3    &   56.8    &   25.2    &   17.7    \\  
0.5 &   $rms^{GLN}$    &   0.4 &   2.2 &   6.0 &   5.4 &   3.6 &   0.9 &   3.7 &   5.7 &   6.8 \\  
    &   $rms^{ODP}$   &   2.1 &   2.3 &   6.4 &   5.9 &   4.7 &   2.7 &   3.8 &   6.3 &   7.5 \\  
    &   $rms^{BS}$  &   2.7 &   3.1 &   11.3    &   10.3    &   7.6 &   2.7 &   25.1    &   9.3 &   5.7 \\  \hline

\end{tabular}
\caption{Simulation performance of distribution forecasts for the data used in Taylor \& Ashe (1983)
Results. The study is based on $10^5$ repetitions, and for each simulated upper triangle, the bootstrap is based on 999 simulations.}
\label{tab:sim:forecast_TaylorAshe}
\end{table}

The second panel of 
Table \ref{tab:sim:forecast}
shows the performance of the traditional chain-ladder. Since the data are log-normal
we expect the chain-ladder to perform poorly. We apply the asymptotic theory of 
Harnau and Nielsen (2017)
and the bootstrap of
England and Verrall (1999)
and
England (2002)
as implemented by Gesmann et al.\ (2015)
The results are generated as before with the difference that the point forecasts are
based on the traditional chain-ladder, while the data remain log-normal.
The actual forecast errors, $true^{ODP}$
are similar to the previous actual errors $true^{GLN}$, particular in the right tail of
the distribution.
The asymptotic distribution approximation, $t^{ODP}$,
and the bootstrap approximation, $BS$, do not provide the same quality of approximations as
$t^{GLN}$ did for $true^{GLN}$.
For large $v=2$ the bootstrap is very poor, possibly because of resampling of large
residuals arising from the mis-specification.

We also simulate the root mean square forecast error for the three methods.
For the log-normal asymptotic distribution approximation this is computed as follows.
We first find mean, standard deviation and quantiles
of the infeasible reserve based on the draws of the lower
triangle $\mathcal{J}$.
This is the true forecast distribution.
For each draw of the upper triangle
$\mathcal{I}$
we then compute
mean, standard deviation and quantiles
of the asymptotic distribution forecast
(\ref{LN_distribution_forecast})
and subtract the
mean, standard deviation and quantiles, respectively,
of the true forecast distribution.
We square, take average across the draws of the upper triangle
$\mathcal{I}$, and then the take the square root.
Similar calculations are done for the over-dispersed approximation and the bootstrap.

The third panel of 
Table \ref{tab:sim:forecast}
shows the root mean square forecast errors.
We see that the generalized log-normal distribution approximation is superior
in all cases and that the bootstrap can be very poor if $v$ is not small. 

In
Table \ref{tab:sim:forecast_TaylorAshe}
we repeat the simulation exercise for the
Taylor and Ashe (1983) data.
For these data we repeated the empirical exercise of
\S\ref{s:emp}, although we do not report the results here.
We found that the generalized log-normal chain-ladder and the over-dispersed chain-ladder
appear to give equally good fit, so that we will expect less difference between the methods
in this case.  We suspect that this arises because of two features in the data.
The Taylor and Ashe triangle has a smaller dimension of $k=10$
and there is less difference between the accident year parameters, see also
Harnau and Nielsen (2017, Table 2).
As before we simulate a log-normal distribution with parameters equal to the
estimates from the data.

Table \ref{tab:sim:forecast_TaylorAshe}
shows that the three methods perform similarly.
In this discussion we focus on the
root mean square error for the 99.5\% quantile which is perhaps of most practical interest.
For large $v=2$ and $v=1$ the
over-dispersed Poisson method actually dominates the generalized log-normal model
even though the data are generated to be log-normal.
For a smaller $v=1/2$ the asymptotic approximation for the
generalized log-normal beats that of the over-dispersed model slightly.
However, the bootstrap appears to be best for $v=1$ and $v=1/2$.

%%%%%%%%%%%%%%%%%%%%%%%%%%%%%%%%%%%%%%%%%%%%%%
%%%%%%%%%%%%%%%%%%%%%%%%%%%%%%%%%%%%%%%%%%%%%%
%%%%%%%%%%%%%%%%%%%%%%%%%%%%%%%%%%%%%%%%%%%%%%
\section{Conclusion}
\label{s:conclusion}

We have presented a new method for distribution forecasting of general insurance reserves in terms of the generalized log-normal model. The forecasts are done under the asymptotic framework which allows users to draw inferences and make model selections easily. This gives an alternative to the traditional chain-ladder where we have the commonly used bootstrap method developed by England and Verrall (1999) and England (2002) along with the recent asymptotic theory of Harnau and Nielsen (2017).

The actuary will have to choose whether the traditional or the normal chain ladder or a third method should be used for a given reserving triangle. In some situations the normal chain ladder will be better than the traditional chain latter as shown in our empirical data analysis and simulation study. In addition, we have considered a number of London market datasets. We compared the standard error over mean forecast trends by year of account with the actuaries' selected volatilities and found that the generalized log-normal trends are more in line with the actuaries selected trends than the over-dispersed Poisson model.

The generalized log-normal model distribution forecasts developed here could also improve the actuarial process for a corporation. The log-normal is also often used in simulating attritional reserve risk for capital modelling. At present this is some times combined with the bootstrap method for the traditional chain ladder. This can result in inconsistencies often between reserving and capital modelling.

A limitation of the log-normal model is that it only fits positive incremental values, while in real life some values can be negative due to reinsurance recoveries, salvage or other data issues such as mis-allocation between classes of business or currencies. In these cases judgements are required and further research must look at how to provide statistical tools to overcome such a limitation.

There is also scope to develop a more advanced model selection process than the model specification tests discussed here. This will give actuaries a statistical basis to select one model over another rather than just eye-balling a distribution fit on a graph. Testing constancy of the dispersion as presented here for the log normal chain ladder and by Harnau (2017) for the traditional chain ladder is a beginning of that research agenda.

The bootstrap method has become popular in recent decades. This is because it usually produces distributions that appear reasonable  and it is a simulation based technique which is favoured by many actuaries. A deeper understanding of the bootstrap method can be developed so that it allows model selections and extensions to generate reserve forecasts under other distributions than the over-dispersed Poisson.

\newpage
\appendix
\section{Appendix: Proofs of Theorems}
\label{s:appendix}

%%%%%
\noindent
\textbf{Proof of Theorem \ref{t:log_normal_distribution}.}
Recall the following results.
A log-normally distributed variable $Y_{ij}$ is positive, hence non-negative.
It is infinitely divisible as shown by
Thorin (1977).
The first three cumulants are
\begin{eqnarray}
    \mathsf{E}(Y_{ij})
    =
    \exp (\mu_{ij}+\omega^2/2)
    ,
\\
    \mathsf{Var}(Y_{ij})
    =
    \exp (2\mu_{ij}+\omega^2)
    \{
        \exp (\omega^2) - 1
    \}
    ,
\\
    \frac{\mathsf{E}\{ Y_{ij} - \mathsf{E} (Y_{ij} ) \}^3}
         {\{ \mathsf{Var}(Y_{ij}) \}^{3/2}}
    =
    \{ \exp (\omega^2) -2 \}^{1/2} 
    \{ \exp (\omega^2) +2 \}
    ,
\end{eqnarray}
see
Johnson, Kotz and Balakrishnan (1994, 
equations 14.8a, 14.8b and 14.9a).

The log-normal distribution is a non-degenerate and non-negative divisible distribution, see Thorin (1977) and
\begin{eqnarray*}
skew(Y) &=& \frac{E(Y-E(Y))^3}{\sqrt{Var(Y)}^3} = \frac{\exp(3\omega^2 )- 3\exp(\omega^2) + 2}{\left(exp(\omega^2)-1\right)^{3/2}}\\
&=& \frac{1 + 3\omega^2 + \frac{1}{2}9\omega^4 - 3\left(1 + \omega^2 + \frac{\omega^4}{2}\right) + 2 + O(\omega^6)}{(1 + \omega^2 -1)^{3/2}}\\
&=& \frac{\left(\frac{9}{2} - \frac{3}{2}\right)\omega^4 + O(\omega^6)}{\omega^3} = 3\omega + O(\omega^3) \to 0.
\end{eqnarray*}
as $\omega\to0$.
Theorem \ref{t:log_normal_distribution} follows by Theorem \ref{t:HN:CLT}, or Theorem 1 in Harnau \& Nielsen (2017).
\hfill$\square$
\bigskip

%%%%%%%%
The next results require the delta method given as follows.
\begin{lemma}
    \label{l:delta_method}
    \textbf{The delta method
        (van der Vaart, 1998, Theorem 3.1)}
    Let $T_\omega$ be a sequence of random vectors or variables indexed by
    $\omega$.
    Suppose
    $\omega^{-1}(T_\omega - \theta)$
    is asymptotically normal
    $\mathsf{N} (0,\Omega )$    
    for $\omega\to 0$
    and that
    $g$
    is a vector or scale valued function
    that is differentiable in a neighbourhood of $\theta$ with derivative $\dot g$.
    Then
    $\omega^{-1}\{g(T_\omega ) - g (\theta)\}$
    is asymptotically normal
    with mean zero and variance
    $\{\dot g(\theta)\} \Omega \{\dot g(\theta)\}'$
    .
\end{lemma}

%%%%%
\noindent
\textbf{Proof of Theorem \ref{t:asymp_logY}.}
Throughout the proof we ignore the indices $i,j$.

1. We show that
\begin{eqnarray}
        \omega^{-1}\{Y - \exp(\mu)\}
    =   \omega^{-1}\{Y - \mathsf{E}(Y)\}
    +   \mathrm{O}(\omega)
\label{Th1:Top}
\end{eqnarray}
First, we add and subtracting $\mathsf{E}(Y)$ term in $Y - \exp(Y)$ to get
\begin{eqnarray}
        \omega^{-1}     \{Y - \exp(Y)\}
    =   \omega^{-1}     \{Y - \mathsf{E}(Y)\}
    +   \omega^{-1}     \{\mathsf{E}(Y) - \exp(\mu)\}
    .
\end{eqnarray}
By
Assumption \ref{as:HN:inf_div:LN}$(i)$
then
$\mathsf{E}(Y) = \exp(\mu + \omega^2/2)$ so that the second term becomes
$$
    \mathcal{E}_2
    =
        \omega^{-1}     \{\mathsf{E}(Y) - \exp(\mu)\}
    =
        \omega^{-1}     \exp(\mu)
        \{\exp(\omega^2 / 2) - 1\}
    .    
$$
Taylor expand the exponential function as
$\exp(\omega^2 / 2) - 1 = \omega^2/2 + \mathrm{O} (\omega^4) $
to get
$$
    \mathcal{E}_2
    =
    \exp(\mu)
    \{
        \omega/2 + \mathrm{O} (\omega^3)
    \}
    =
    \mathrm{O}(\omega)
    ,
$$
since the canonical parameter $\xi$ is fixed, and hence $\mu_{ij}$
is fixed.
The expression
(\ref{Th1:Top})
then follows.

2. We show that
\begin{eqnarray}
    \omega^{-1}
    \{ Y \exp(-\mu) -1 \}
%    =
%    \frac{Y- \exp(\mu)}{\omega\exp(\mu)}
    \xrightarrow{\mathsf{D}} \mathsf{N}(0,1)
    .
\label{Th1:2}
\end{eqnarray}
Apply
(\ref{Th1:Top})
and divide by $\exp(\mu)$,
multiply and divide by $\surd\mathsf{Var}(Y)/\omega$ and
$\mathsf{E}(Y)$ to get
\begin{eqnarray*}
    \frac{Y - \exp(\mu)}{\omega \exp(\mu)}
    &=&
    \frac{Y - \mathsf{E}(Y)}{\omega \exp(\mu)}
    +
    \mathrm{O}(\omega)
    =
    \{\frac{Y - \mathsf{E}(Y)}{\surd\mathsf{Var}(Y)}    \}
    \{\frac{\surd\mathsf{Var}(Y)}{\omega \mathsf{E}(Y)} \}
    \{\frac{\mathsf{E}(Y)}{\exp(\mu)}                   \}
    +
    \mathrm{O}(\omega)
    .
\end{eqnarray*}
Assumption \ref{as:HN:inf_div:LN}$(i,iii)$
implies that the second and third terms converge to unity.
Theorem \ref{t:HN:CLT},
using Assumption \ref{as:HN:inf_div}.
shows the first term is asymptotically normal.
Dividing by $\exp(\mu)$
in numerator and denominator establishes
(\ref{Th1:2}).

3. Apply the delta method in
Lemma \ref{l:delta_method}
to
(\ref{Th1:2})
with
$T_\omega=Y\exp (-\mu)$
and
$\theta= 1$
and choose
$g(t) = \log(t)+\mu$,
so $\dot g(t) = 1/t$.
Then
$g(T_\omega)=\log Y$
and
$g(\theta)=\mu$
while
$\dot g (\theta) = 1$
so that
$\omega^{-1} (\log Y - \mu )$
is asymptotically standard normal as desired.
\hfill$\square$
\bigskip

%%%%%%%%%%%%%%%%%%
\noindent
\textbf{Proof of Theorem \ref{t:asymp_Y}.}
    Theorem \ref{t:HN:CLT}
    shows that
    $\{Y_{ij}-\mathsf{E} (Y_{ij})\}/\surd \mathsf{Var} (Y_{ij})$
    is asymptotically standard normal.
    Now, 
    Assumption \ref{as:HN:inf_div:LN}$(iii)$
    shows
    $\mathsf{Var}(Y_{ij})/\{ \omega^2\mathsf{E}^2(Y_{ij})\}\to 1$,
    while 
    Assumption \ref{as:HN:inf_div:LN}$(i,ii)$
    implies
    $\log \mathsf{E}(Y_{ij})\to \mu_{ij}$.
    Combine these three results to get the desired statement.
\hfill$\square$
\bigskip

%%%%%%%%%%
\noindent
\textbf{Proof of Theorem \ref{t:asymp_est}.}
The model equation is
$y_{ij} = \log Y_{ij} = X_{ij}'\xi + \varepsilon_{ij}$,
see
(\ref{model_logN}).
Theorem \ref{t:asymp_logY},
using
Assumptions \ref{as:HN:inf_div}, \ref{as:HN:inf_div:LN},
shows that the vector of innovations 
$\omega^{-1} \varepsilon = \omega^{-1} ( y - X\xi )$
is asymptotically standard normal as
$\omega\to 0$.
We can then use standard least squares distribution theory in the limit. 

Recall $\hat\xi = (X^{'}X)^{-1}X^{'} y$, see
(\ref{est_logN:xi}).
Substitute $y=X\xi + u$ to get
\begin{eqnarray*}
    \omega^{-1}
    (\hat\xi - \xi)
    &=&
    \omega^{-1}
    \{(X^{'}X)^{-1}X^{'}(X\xi + \varepsilon) - \xi
    \}
    = (X^{'}X)^{-1}X^{'}
    (\omega^{-1}
    \varepsilon
    )
    .
\end{eqnarray*}
Since
$\omega^{-1}\varepsilon \xrightarrow{D}
\mathsf{N}(0,I_n)$,
we have
$
    (\omega^{-1}
    (\hat\xi - \xi)  \xrightarrow{D}
    \mathsf{N}
    \{ 0,  (X^{'}X)^{-1}\}
$
as required.

The residuals in 
(\ref{est_logN:xi})
can be written as
$\hat\varepsilon=
    P_\perp
    y
$,
where
$
    P_\perp
    =
    \{I_n - X(X^{'}X)^{-1}X^{'}\}
$
is an orthogonal projection matrix so that
$P_\perp=P_\perp'$
and
$P_\perp^2=P_\perp$.
Inserting the model equation this becomes
$\hat\varepsilon=
    P_\perp
    \varepsilon
    ,
$
while
$P_\perp X=0$.
Since
$\omega^{-1}\varepsilon \xrightarrow{D}
\mathsf{N}(0,I_p)$,
then
$\omega^{-1}P_\perp\varepsilon \xrightarrow{D}
\mathsf{N}(0,P_\perp)$,
so that
$
    \omega^{-2} s^2    
$
is asymptotically
$   \chi^2_{n-p}/(n-p)
$
noting $\mathrm{tr}(P_\perp)=n-p$.

Finally $\hat\xi$ and $s^2$
are asymptotically independent, since
$\hat\xi-\xi$ and
$s^2$
are functions of $X'\varepsilon$
and
$P_\perp \varepsilon$,
while
$\omega^{-1}\varepsilon$
is asymptotically standard normal, while
$P_\perp X=0$.
\hfill$\square$
\bigskip

%%%%%%%%%%%%%%%%
\noindent
\textbf{Proof of Theorem \ref{t:asymp_forecast}.}
Recall the forecast taxonomy
(\ref{forecast:logN:taxonomy}),
summed over $\mathcal{A}$.

The first contribution is the process error and satisfies
\begin{equation*}
    \omega^{-1}
    \{Y_\mathcal{A} - E(Y_\mathcal{A})\}
    =
    \omega^{-1}
    \sum_{i,j\in\mathcal{A}}
    \{Y_{ij} - E(Y_{ij})\}
    .
\end{equation*}
This is a sum of independent terms, each of which is asymptotically
$\mathsf{N} \{ 0, \exp (2\mu_{ij}) \}$
by
Theorem \ref{t:asymp_Y}.
Therefore, 
$    \omega^{-1}
    \{Y_\mathcal{A} - E(Y_\mathcal{A})\}
$
is asymptotically
$\mathsf{N} ( 0, \varsigma^2_{\mathcal{A},process} )$,
where
$
    \varsigma^2_{\mathcal{A},process}
    = 
    \sum_{i,j\in\mathcal{A}}
    \exp (2\mu_{ij})
$
as stated in
(\ref{forecast:error_process}),
(\ref{forecast:var_process}).

The second contribution is the estimation error from
$\hat\xi$.
Theorem \ref{t:asymp_est}
shows that as
$\omega\to 0$
then
$\omega^{-1}(\hat\xi-\xi)\xrightarrow{D} \mathsf{N} \{ 0, (X^{'}X)^{-1}\}$.
%so that
%$\omega^{-1}(X\hat\xi-X\xi)\xrightarrow{D} \mathsf{N} \{ 0, X(X^{'}X)^{-1}X'\}$.
Apply the delta method in
Lemma \ref{l:delta_method}
with
$T_\omega=\hat\xi$
and 
$g(T)=
\sum_{i,j\in\mathcal{J}}
\exp (X_{ij}'\xi )$,
so that
$\dot g(T)=
\sum_{i,j\in\mathcal{J}}
\exp (X_{ij}'\xi ) X_{ij}'$.
Therefore, 
$    \omega^{-1}
    \{
        \exp (X_{ij}'\hat\xi )
        -
        \exp (X_{ij}'\xi )
    \}
$
is asymptotically
$\mathsf{N} ( 0, \varsigma^2_{\mathcal{A},estimation} )$,
where
$
    \varsigma^2_{\mathcal{A},estimation}
$    
is given in
(\ref{forecast:var_estimation}).
Further, by continuity
$\exp(\omega^2/2)\to 1$
as
$\omega^2\to 0$.
In combination we arrive at 
(\ref{forecast:error_estimation}).

The third term is the contribution from estimation error of $s^2$.
By continuity, we get
$\exp(\omega^2/2)\to 1$
as
$\omega^2\to 0$,
while
$\sum_{i,j\in\mathcal{A}}\exp(X_{ij}'\xi)$
is fixed.
Rewrite
$
    s^2
    =
    ( s^2/\omega^2 ) \omega^2 
$.
Since
$s^2/\omega^2$
converges in distribution by
Theorem \ref{t:asymp_est}
as
$\omega^2\to 0$
then
$s^2$
vanishes in probability.
Applying the exponential function, which is a continuous mapping,
yields that
$\exp(s^2/2)\to 1$ in probability
and so does the entire third term.

The process error and the estimation error are independent as they are based on
the independent upper and lower triangles
$\mathcal{J}$
and
$\mathcal{I}$.
Therefore, the first and second contributions to
the forecast taxonomy
(\ref{forecast:logN:taxonomy})
are independent, while the third contribution vanishes, so that
\begin{equation*}
    \omega^{-1}
    \{Y_\mathcal{A} - E(Y_\mathcal{A})\}
    \overset{\mathsf{D}}{\to}
    \mathcal{N} (
    \varsigma^2_{\mathcal{A},process}+
    \varsigma^2_{\mathcal{A},estimation}
    )
    ,
\end{equation*}
which is asymptotically independent of $s^2$.
Further,
$s^2/\omega^2$
is asymptotically
$\chi^2_{n-p}/(n-p)$
so that
$
    s^{-1}
    \{Y_\mathcal{A} - E(Y_\mathcal{A})\}
$
is asymptotically
$\mathsf{t}_{n-p}$
as desired.
\hfill$\square$

%%%%%%%%%%%%%%%%%%%%%%%%%%%%%%%%%%%%%%%%%%%%%
\newpage
\section{Further table}
\label{s:tables}

\begin{table}[ht]
\centering
\begin{tabular}{lrrlrr}
                                                                                                            \hline
    $\mu_{11}$              &  7.689& $\mu_{21}-\mu_{11}$   &  0.0929& $\mu_{12}-\mu_{11}$   &  2.076    \\
    $\Delta^2\alpha_3$      & -0.133& $\Delta^2\beta_3$     & -1.347& $\Delta^2\gamma_3$    & 0.343    \\
    $\Delta^2\alpha_4$      & -0.422& $\Delta^2\beta_4$     & -0.690& $\Delta^2\gamma_4$    & 0.044    \\
    $\Delta^2\alpha_5$      &  0.427& $\Delta^2\beta_5$     & -0.134& $\Delta^2\gamma_5$    & -0.312    \\
    $\Delta^2\alpha_6$      & -0.532& $\Delta^2\beta_6$     & -0.272& $\Delta^2\gamma_6$    & 0.170    \\
    $\Delta^2\alpha_7$      &  0.181& $\Delta^2\beta_7$     &  0.036& $\Delta^2\gamma_7$    &  -0.253    \\
    $\Delta^2\alpha_8$      &  0.177& $\Delta^2\beta_8$     & -0.297& $\Delta^2\gamma_8$    & 0.249   \\
    $\Delta^2\alpha_9$      &  0.008& $\Delta^2\beta_9$     &  0.131& $\Delta^2\gamma_9$    &  0.065    \\
    $\Delta^2\alpha_{10}$   & -0.118& $\Delta^2\beta_{10}$  & -0.090& $\Delta^2\gamma_{10}$ & -0.042    \\
    $\Delta^2\alpha_{11}$   &  0.119& $\Delta^2\beta_{11}$  &  0.219& $\Delta^2\gamma_{11}$ &  -0.268   \\
    $\Delta^2\alpha_{12}$   & -0.471& $\Delta^2\beta_{12}$  & -0.073& $\Delta^2\gamma_{12}$ & 0.335    \\
    $\Delta^2\alpha_{13}$   &  0.050& $\Delta^2\beta_{13}$  & -0.003& $\Delta^2\gamma_{13}$ & -0.341    \\
    $\Delta^2\alpha_{14}$   &  0.707& $\Delta^2\beta_{14}$  & -0.321& $\Delta^2\gamma_{14}$ & 0.247    \\
    $\Delta^2\alpha_{15}$   &  0.018& $\Delta^2\beta_{15}$  &  0.255& $\Delta^2\gamma_{15}$ &  -0.010    \\
    $\Delta^2\alpha_{16}$   & -0.579& $\Delta^2\beta_{16}$  &  0.709& $\Delta^2\gamma_{16}$ &  0.095    \\
    $\Delta^2\alpha_{17}$   &  0.436& $\Delta^2\beta_{17}$  & -1.276& $\Delta^2\gamma_{17}$ & -0.227    \\
    $\Delta^2\alpha_{18}$   &  0.031& $\Delta^2\beta_{18}$  &  0.984& $\Delta^2\gamma_{18}$ &  0.202    \\
    $\Delta^2\alpha_{19}$   & -0.258& $\Delta^2\beta_{19}$  & -0.463& $\Delta^2\gamma_{19}$ & 0.229    \\
    $\Delta^2\alpha_{20}$   &  0.890& $\Delta^2\beta_{20}$  &  0.034& $\Delta^2\gamma_{20}$ &  0.236    \\
    $s^2$                   &  0.181& $RSS$                 & 27.626&                       &           \\  \hline    
\end{tabular}
\caption{Estimates for the US casualty data for extended chain-ladder, $\mathsf{H}_{apc}$.}
\label{tab:XL:LNestimates:apc}
\end{table}

%%%%%%%%%%%%%%%%%%%%%%%%%%%%%%%%%%%%%%%%%%%%%
%%%%%%%%%%%%%%%%%%%%%%%%%%%%%%%%%%%%%%%%%%%%%
%%%%%%%%%%%%%%%%%%%%%%%%%%%%%%%%%%%%%%%%%%%%%

\newpage
\section{References}
\begin{description}
        \setlength\itemsep{-5pt}
        \setlength\labelsep{0pt}
\item
    Barnett, G. and Zehnwirth, B. \ (2000)
    Best estimates for reserves. 
   \textit{Proceedings of the Casualty Actuarial Society} 87, 245-321.
\item
    Bartlett, M.S.\ (1937)
    Properties of sufficiency and statistical tests.
    \textit{Proceedings of the Royal Society of London} series A 160, 268--282.
\item
    Beard, R.E.\, Pentik{\" a}inen, T.\ and Pesonen, E.\ (1984)
    \textit{Risk Theory}, 3rd edn.
    London: Chapman \& Hall.
\item
    Doray, L.G.\ (1996)
    UMVUE of the IBNR reserve in a lognormal linear regression model.
    \textit{Insurance: Mathematics and Economics} 18, 43--57.
\item
    England, P.\ (2002)
    Addendum to "Analytic and bootstrap estimates of prediction errors in claims reserving".
    \textit{Insurance: Mathematics and Economics} 31, 461--466.
\item
    England, P.\ and Verrall, R.\ (1999)
    Analytic and bootstrap estimates of prediction errors in claims reserving.
    \textit{Insurance: Mathematics and Economics} 25, 281--293.
\item
    England, P.D.\ and Verrall, R.J. (2002)
    Stochastic claims reserving in
general insurance. \textit{British Actuarial Journal} 8, 519--44.
\item
    Gesmann, M., Murphy, D., Zhang, Y., Carrato, A., Crupi, G., W{\" u}thrich, M.\ and Concina, F.\ (2015)
    Chainladder: Statistical methods and models for claims reserving in general insurance.
    \verb|cran.R-project.org/package=ChainLadder|.
\item
    Harnau, J.\ (2017)
    Misspecification tests for chain-ladder models.
    Department of Economics, University of Oxford, Discussion Paper 840. 
\item
    Harnau, J.\ and Nielsen, B.\ (2017)
    Over-dispersed age-period-cohort models.
    Nuffield College Discussion Paper.
    To appear in
    \textit{Journal of the American Statistical Association}.
\item
    Hendry, D.F.\ and Nielsen, B.\ (2007)
    \textit{Econometric Modeling}.
    Princeton, NJ: Princeton University Press.
\item
    Hertig, J.\ (1985)
    A statistical approach to {IBNR}-reserves in marine reinsurance.
    \textit{ASTIN Bulletin} 15, 171--183.    
\item
    Johnson, N.~L., Kotz, S.\ and Balakrishnan, N.\ (1994)
    \textit{Continuous Univariate Distributions}
    volume 1,
    2nd edn.
    New York: Wiley.
\item
    Kremer, E.\ (1982)
    IBNR-Claims and the Two-way model of ANOVA. 
    \textit{Scandinavian Actuarial Journal}, 47--55.
%\item
%    Kremer, E.\ (1985)
%    \textit{Einf\"{u}hrung in die Versicherungsmathematik}.
%    G\"{o}ttingen: Vandenhoeck \& Ruprecht.
\item
    Kuang, D., Nielsen B.\ and Nielsen J.P.\ (2008a)
    Identification of the age-period-cohort model and the extended chain-ladder model.
    \textit{Biometrika} 95, 979--986.
\item
    Kuang, D., Nielsen B.\ and Nielsen J.P.\ (2008b)
    Forecasting with the age-period-cohort model and the extended chain-ladder model.
    \textit{Biometrika} 95, 987--991.
\item
    Kuang, D., Nielsen B.\ and Nielsen J.P.\ (2009)
    chain-ladder as Maximum Likelihood Revisited.
    \textit{Annals of Actuarial Science} 4, 105-121.
\item
    Kuang, D., Nielsen B.\ and Nielsen J.P.\ (2011)
    Forecasting in an extended chain-ladder-type model.
    \textit{Journal of Risk and Insurance} 78, 345--359.
\item
    Kuang, D., Nielsen B.\ and Nielsen J.P.\ (2015)
    The geometric chain-ladder.
    \textit{Scandinavian Actuarial Journal}, 278--300.
%\item
 %    Mack, T. (1991)
 %   A simple parametric model for rating automobile insurance or estimating IBNR claims reserves.
 %    \textit{ASTIN Bulletin}, 21, 93-109.
\item
    Mack, T. (1999)
    The standard error of chain ladder reserve estimates: Recursive calculation and inclusion of a tail factor.
    \textit{ASTIN Bulletin} 29, 361--366.
%\item
 %    Mack, T. \ and Venter, G. (2000) 
 %    A comparison of stochastic models that reproduce the Chain Ladder estimates.
 %    Insurance: Mathematics and Economics 26, 101-107.
\item
    Mart\'{i}nez-Miranda, M.D., Nielsen, B.\ and Nielsen, J.P.\ (2015)
    Inference and forecasting in the age-period-cohort model with unknown exposure with an application to mesothelioma mortality.
    \textit{Journal of the Royal Statistical Society} series A 178, 29--55.
\item
    Nielsen, B.\ (2015)
    apc: An R package for age-period-cohort analysis.
    \textit{R Journal} 7, 52--64.
\item
    R Core Team (2017)
    R: A language and environment for statistical computing. \\
    \verb|www.R-project.org|.
\item
    Renshaw, A.E.\ (1989)
    Chain ladder and interactive modelling (Claims reserving and GLIM).
    \textit{Journal of the Institute of Actuaries} 116, 559--587.
%\item
%   Renshaw A. E. and Verrall R. J. (1998). 
%   A stochastic model underlying the chainladder technique. 
%   \textit{British Actuarial Journal} 4, 903923.
\item
Taylor, G. ~C. (1979) 
Statistical Testing of a Non-Life Insurance Model. 
Proceedings Actuarial Sciences Institute, Act. Wetemschappen, Katholieke Univ. Leuven, Belgium.
\item
    Taylor, G.~C. and Ashe, F.~R. (1983)
    Second Moments of Estimates of Outstanding Claims.
    \textit{Journal of Econometrics} 23, 37--61.
\item
    Thorin, O.\ (1977)
    On the infinite divisibility of the lognormal distribution.
    \textit{Scandinavian Actuarial Journal} 1977, 121--148.
\item
    van der Vaart, A.W.\ (1998)
    \textit{Asymptotic Statistics}
    Cambridge: Cambridge University Press.
\item
    Verrall, R.J. (1991)
    On the estimation of reserves from log-linear models.
    \textit{Insurance: Mathematics and Economics} 10, 75--80.
\item
    Verrall, R.J. (1994)
    Statistical methods for the chain-ladder technique.
    \textit{Casualty Actuarial Society Forum, Spring 1994}, 393--446.
\item
    XL Group (2017)
    \textit{2016 Global Loss Triangles} \\
    {\footnotesize
    \verb|http://phx.corporate-ir.net/phoenix.zhtml?c=73041&p=irol-financialreports|
    }    
\item
    Zehnwirth, B.\ (1994)
    Probabilistic development factors with applications to loss reserve variability, prediction intervals, and risk based capital.
    \textit{Casualty Actuarial Society Forum, Spring 1994}, 447--605.
\end{description}

\end{document}